\documentclass[preprint,11pt,nopreprintline,numbers]{elsarticle}

\usepackage[T1]{fontenc}
\usepackage[utf8]{inputenc}
\usepackage{lmodern}
\usepackage{amssymb,amsmath,mathtools}
\usepackage{xcolor}
\usepackage{graphicx,booktabs,array,tabularx}
\usepackage{placeins,microtype,flafter,needspace}
\usepackage[a4paper,margin=25mm]{geometry}
\usepackage[hidelinks]{hyperref}
\hypersetup{pdfauthor={John Rick Manzanares, Richard Leslie Abel, Pawel Dlotko},
pdftitle={Frame-invariant topological representations of trabecular bone microarchitecture for strength prediction}}
\graphicspath{{./}}
\newcolumntype{L}[1]{>{\raggedright\arraybackslash}p{#1}}
\newcolumntype{Y}{>{\raggedright\arraybackslash}X}
\newtheorem{definition}{Definition}
\newtheorem{proposition}{Proposition}
\newproof{proof}{Proof}
\newcommand{\R}{\mathbb R}
\newcommand{\bonefrac}{\mathrm{BV/TV}}

\colorlet{revisioncolor}{black}
\newenvironment{revision}{\begingroup\color{revisioncolor}}{\endgroup}
\journal{Computer Methods and Programs in Biomedicine}

\title{Frame-invariant topological representations of trabecular bone microarchitecture for strength prediction}
\author[1,2]{John Rick Manzanares\corref{cor1}}
\ead{jdolormanzanares@impan.pl}
\author[3]{Richard Leslie Abel}
\ead{richard.abel@imperial.ac.uk}
\author[1]{Pawe\l{} D\l{}otko}
\ead{pawel.dlotko@impan.pl}
\cortext[cor1]{Corresponding author. Email: \nolinkurl{jdolormanzanares@impan.pl}}
\affiliation[1]{addressline={Institute of Mathematics of the Polish Academy of Sciences},city={Warsaw},country={Poland}}
\affiliation[2]{organization={International Environmental Doctoral School},
addressline={University of Silesia in Katowice},city={Sosnowiec},country={Poland}}
\affiliation[3]{organization={Department of Surgery and Cancer, Faculty of Medicine},
addressline={Imperial College London},city={London},country={United Kingdom}}

\begin{document}

\begin{abstract}
\indent
\textbf{Background and Objectives:} Directional topological representations of trabecular bone should retain interpretable structural information without depending on an arbitrary transverse coordinate frame. We develop a frame-invariant directional filtration and compare its strength prediction with signed distance persistent homology and conventional morphometry.

\textbf{Methods:} Twenty-four human trabecular cores were analyzed using persistence images, directional Betti tensors, and nested ridge regression over 13 validation groups. The directional construction combines cone occupancy, directional covariance, and principal axis degeneracy. Equal bone removal experiments and a pair closely matched in morphometry were used to examine whether topological differences tracked changes in simulated elastic stiffness.

\textbf{Results:} The original combined persistence image model had a root mean squared error (RMSE) of 1.952~MPa, compared with 2.001~MPa for morphometry; the paired difference was $-0.049$~MPa with a 95\% bootstrap interval of $[-0.491,0.393]$~MPa. Exploratory signed distance persistent homology in dimension zero gave an RMSE of 1.639~MPa. The post hoc frame-invariant directional dimension zero model gave 1.610~MPa, compared with 1.877~MPa for dimension one and 1.630~MPa for a harmonized signed distance dimension zero model. The paired RMSE difference between the invariant and signed distance models was $-0.021$~MPa with a 95\% interval of $[-0.234,0.210]$~MPa. Localized removal reduced stiffness more than diffuse removal in 19 of 24 cores despite producing smaller $H_0$ persistence image changes.

\textbf{Conclusions:} Frame invariance removes the dependence of directional topology on an arbitrary transverse coordinate frame. Connected component representations warrant external evaluation for strength prediction, while the mechanical experiments limit their interpretation as scalar stiffness surrogates.
\end{abstract}


\maketitle

\section{Introduction}

A quantitative descriptor of trabecular bone is a map from a three-dimensional structure to a lower-dimensional representation. In conventional morphometry, this reduction is achieved by replacing the full trabecular architecture with a small number of scalar quantities that summarize particular geometric properties of the structure. The simplest example is the bone volume fraction, which records the proportion of the region of interest occupied by bone. Trabecular thickness and separation summarize characteristic local length scales, while connectivity density and degree of anisotropy describe aspects of network organization and directional structure. These descriptors are useful because they deliberately discard most of the image while retaining quantities with established geometric interpretations \cite{https://doi.org/10.1111/j.1600-0501.2004.00969.x,BoneJ2021,ODGAARD1993173,ODGAARD1997315}. The resulting compression leaves a mathematical question: what information discarded by conventional morphometry is retained by a multiscale description of the full segmented microarchitecture, and is any such information associated with mechanical strength \cite{Gomberg2000,Pothuaud2002}?

Persistent homology addresses the first part of this question by studying topology across a nested family of spaces rather than at a single binary representation \cite{Zomorodian2004,CohenSteiner2006}. Such a nested family is called a filtration. Let $B\subset R$ denote the segmented bone phase inside a region of interest $R$, and define the signed Euclidean distance function $s:R\rightarrow\mathbb{R}$ by assigning negative distance to voxels inside bone and positive distance to voxels in marrow. For $t\in\mathbb{R}$, the corresponding sublevel sets
\[
B_t=\{x\in R:s(x)\leq t\}
\]
satisfy $B_u\subseteq B_t$ whenever $u\leq t$, and therefore form a \emph{filtration}.

At a filtration value $t$, the zeroth homology group $H_0(B_t)$ describes connected components, the first homology group $H_1(B_t)$ describes independent one-dimensional cycles, and the second homology group $H_2(B_t)$ describes independent two-dimensional cycles, which in three dimensions can surround enclosed voids.

Because $s$ is measured in physical distance, the filtration parameter $t$ is measured in micrometers. For $t=-r<0$, the set $B_t$ contains bone lying at least $r$ micrometers from marrow; at $t=0$, it recovers the segmented bone; and for $t>0$, it additionally includes marrow lying within distance $t$ of bone. Birth and death values in persistent homology therefore identify both when a topological feature appears or disappears and the physical spatial scale at which that change occurs \cite{Pritchard2023,SONG2025102857}.

The three homology dimensions describe different properties of the evolving filtered structure and should not be treated as interchangeable summaries. In the signed distance filtration, $H_0$ records the appearance and merging of connected regions as progressively less interior bone is admitted into the filtration. By contrast, classes in $H_1$ and $H_2$ arise from cycles and enclosed regions of the evolving filtered complex. These features are defined by the filtration itself and need not correspond directly to anatomical tunnels or cavities present in the original segmented bone at $t=0$.

This interpretation gives dimension zero a direct geometric relation to the internal scale and connectivity of the trabecular structure. As the filtration threshold increases from negative values toward zero, connected regions lying deeper within bone appear and subsequently merge with one another. The resulting persistence therefore depends both on the depth of interior bone regions and on the spatial scales at which those regions become connected. This information is related to, but not equivalent to, trabecular thickness. Local thickness is defined from the diameters of maximal inscribed spheres, whereas dimension zero persistence describes connectivity changes across a nested family of interior bone regions.

The original predictive analysis treated the combined information from $H_0$, $H_1$, and $H_2$ as the primary topological representation and compared this model with conventional morphometry. Models based on the individual homology dimensions were evaluated separately as exploratory analyses. Among these, $H_0$ produced the lowest observed prediction error and, independently, admits the direct geometric interpretation described above. We therefore emphasize $H_0$ when interpreting the results for the individual homology dimensions, while retaining the combined representation as the original primary benchmark. The observed performance of $H_0$ remains exploratory and requires evaluation in an independent cohort.

The signed distance filtration describes spatial scale but does not explicitly distinguish one physical direction from another. We therefore performed a secondary analysis asking whether directional topology contains additional information about trabecular architecture. In reproducing a directional topological construction proposed in a recent preprint \cite{Topolnicki2026Directional}, we found that its representation could depend on an arbitrary choice of coordinate frame perpendicular to the direction being studied. This motivated a frame-invariant directional filtration whose coordinates depend only on quantities intrinsic to the selected physical direction and the local trabecular geometry. Because this construction was developed after the dependence on the transverse coordinate frame and the earlier empirical results were known, we treat it as a post hoc methodological extension whose findings require independent evaluation.

Prediction and mechanical interpretation address different claims. A representation may be associated with strength across specimens without implying that a change in that representation must produce a corresponding change in mechanical response. The methodological contribution is a directional cubical filtration that removes this ambiguity in the transverse frame, together with an exact computation of its sampled Betti numbers and a controlled evaluation of its interpretation in human bone. We therefore address three related questions. First, how do the individual homology dimensions compare in their association with measured apparent compressive strength? Second, how does the resulting focal topological representation compare with conventional morphometry and other structural descriptions of trabecular architecture? Third, does the association observed across specimens persist under controlled geometric perturbations designed to separate the amount of bone from its spatial organization? Together, these analyses distinguish the predictive value of a representation from the stronger claim that changes in that representation have a consistent mechanical interpretation.

\begin{revision}
\section{Mathematical framework}
\label{sec:framework}

This section fixes the mathematical objects used throughout the study. First, each segmented image is represented as a finite cubical complex. Cubical homology and filtrations are then introduced before the signed distance and directional constructions.

\subsection{Voxel geometry and cubical complexes}
\label{sec:cubical_preliminaries}

For specimen $i$, let $\Omega_i\subset\mathbb Z^3$ be the finite three-dimensional voxel lattice, let $R_i\subset\Omega_i$ be the region of interest, and let $B_i\subset R_i$ be the segmented bone phase. The isotropic voxel spacing is $h_i>0$. A voxel index $x=(x_1,x_2,x_3)\in\Omega_i$ has physical center $h_i x$, relative to a fixed origin. For a voxel set $A\subseteq\Omega_i$, we write $|A|$ for its number of voxels, so its physical volume is $h_i^3|A|$.

To pass from a voxel array to a topological space, associate to each voxel index $x\in\Omega_i$ the closed physical cube
\begin{equation}
Q_i(x)=h_i x+\left[-\frac{h_i}{2},\frac{h_i}{2}\right]^3.
\label{eq:voxel_cube}
\end{equation}
A \emph{cubical cell} is any face of one of these voxel cubes. Thus, in three dimensions, cubical cells have dimensions $0$, $1$, $2$, or $3$: vertices, edges, square faces, and voxel cubes. If $\sigma$ and $\tau$ are cells, we say that $\tau$ is a \emph{face} of $\sigma$ when $\tau\subseteq\sigma$ is obtained by fixing one or more coordinates of $\sigma$ at an endpoint.

\begin{definition}
A finite \emph{cubical complex} $K$ is a finite collection of cubical cells such that every face of every cell in $K$ also belongs to $K$, and the nonempty intersection of any two cells in $K$ is a face of each. A subset $L\subseteq K$ that is itself closed under faces is a \emph{cubical subcomplex} of $K$.
\end{definition}

For a voxel set $A\subseteq R_i$, define
\begin{equation}
\operatorname{Cub}_i(A)
=
\left\{\sigma:\sigma\text{ is a face of }Q_i(x)\text{ for some }x\in A\right\}.
\label{eq:cub_operator}
\end{equation}
The ambient ROI complex is
\[
K_i^{\mathrm{ROI}}=\operatorname{Cub}_i(R_i),
\]
and the segmented bone at threshold zero will later appear as the subcomplex $\operatorname{Cub}_i(B_i)$. When no ambiguity is possible, we write $\operatorname{Cub}(A)$ instead of $\operatorname{Cub}_i(A)$.

The conventional morphometric representation is defined on the same segmented voxel set. In particular,
\[
f_i=\bonefrac(B_i;R_i)=\frac{|B_i|}{|R_i|},
\]
where the voxel volume $h_i^3$ cancels. The primary morphometric map is
\[
\mathcal M(B_i;R_i)=m_i=(f_i,t_i,s_i,c_i,a_i)\in\R^5,
\]
where $t_i$ and $s_i$ are mean local trabecular thickness and separation, $c_i$ is connectivity density, and $a_i$ is degree of anisotropy. Local thickness is defined from maximal spheres contained in the corresponding phase \cite{Hildebrand1997}; connectivity density is based on the Euler characteristic with an edge correction and volume normalization \cite{ODGAARD1993173}; and degree of anisotropy summarizes directional variation in mean intercept length \cite{ODGAARD1997315}. Numerical conventions are given in \ref{app:images} \cite{BoneJ2021}.

\subsection{Cubical homology and filtrations}
\label{sec:cubical_homology}

Homology is computed over the field $\mathbb F_2=\{0,1\}$, with addition and multiplication modulo two. This choice removes orientation signs from the boundary operator and is standard for complexes built from (binary) images.

For a finite cubical complex $K$, let $C_k(K;\mathbb F_2)$ be the vector space whose basis consists of the $k$-dimensional cells of $K$. Its elements are finite formal sums of $k$-cells with coefficients in $\mathbb F_2$ and are called $k$-\emph{chains}. The cubical boundary map
\[
\partial_k:C_k(K;\mathbb F_2)\longrightarrow C_{k-1}(K;\mathbb F_2)
\]
sends each $k$-cell to the sum of its faces of codimension one. These maps satisfy
\[
\partial_{k-1}\circ\partial_k=0.
\]
Set $Z_k(K)=\ker\partial_k$, the space of cycles, and $B_k(K)=\operatorname{im}\partial_{k+1}$, the space of boundaries. Since $\partial_k\circ\partial_{k+1}=0$, we have $B_k(K)\subseteq Z_k(K)$. The $k$th homology group is the quotient
\begin{equation}
H_k(K;\mathbb F_2)=Z_k(K)/B_k(K),
\label{eq:homology_definition}
\end{equation}
and its dimension
\begin{equation}
\beta_k(K)=\dim_{\mathbb F_2}H_k(K;\mathbb F_2)
\label{eq:betti_definition}
\end{equation}
is the $k$th \emph{Betti number}. In three-dimensional image complexes, $\beta_0$ counts connected components, $\beta_1$ counts independent one-dimensional cycles, and $\beta_2$ counts independent two-dimensional cycles that can surround enclosed regions. These geometric descriptions always refer to the particular complex being analyzed; they need not correspond directly to anatomical tunnels or cavities in the original segmentation.

A \emph{one-parameter filtration} of a complex $K$ is a family of subcomplexes $(K(t))_{t\in\mathbb R}$ such that, if $u\le t$, then $K(u)\subseteq K(t)$. More generally, equip $\mathbb R^m$ with the coordinatewise partial order in which $\mathbf s\preceq\mathbf t$ if and only if $s_j\le t_j$ for every $j$. An $m$-parameter filtration is a family that preserves this order $(K(\mathbf t))_{\mathbf t\in\mathbb R^m}$: if $\mathbf s\preceq\mathbf t$, then $K(\mathbf s)\subseteq K(\mathbf t)$. The signed distance construction below has one parameter, whereas the directional construction has three.

\subsection{Signed Euclidean distance persistence}
\label{sec:sedt_persistence}

For a nonempty voxel set $A\subseteq\Omega_i$, define the center-to-center Euclidean distance
\[
d_i(x,A)=h_i\min_{a\in A}\|x-a\|_2,
\]
where $h_i$ is expressed in micrometers. Assuming that both bone and marrow are nonempty within $R_i$, the \emph{signed Euclidean distance transform} is the map $\psi_i:\Omega_i\rightarrow\mathbb R\cup\{+\infty\}$ defined by
\begin{equation}
\psi_i(x)=
\begin{cases}
-d_i(x,R_i\setminus B_i), & x\in B_i,\\
 d_i(x,B_i), & x\in R_i\setminus B_i,\\
+\infty, & x\notin R_i.
\end{cases}
\label{eq:sedt}
\end{equation}
Negative values measure depth within bone relative to marrow voxel centers, whereas positive values measure the distance of marrow voxels from bone. Voxels outside the ROI are excluded rather than treated as marrow, so the artificial cylindrical crop boundary does not itself enter the distance transform.

For $t\in\mathbb R$, define
\begin{equation}
K_i(t)=\operatorname{Cub}_i\!\left(\{x\in R_i:\psi_i(x)\le t\}\right).
\label{eq:sedt_complex}
\end{equation}
Because a sublevel set can only grow as $t$ increases, $K_i(u)\subseteq K_i(t)$ whenever $u\le t$. Hence $(K_i(t))_{t\in\mathbb R}$ is a one-parameter filtration of $K_i^{\mathrm{ROI}}$. Since $R_i$ is finite, only finitely many distinct complexes occur.

The negative part of the filtration has a direct geometric interpretation. For $r>0$, let
\[
B_i(r)=\{x\in B_i:d_i(x,R_i\setminus B_i)\ge r\}.
\]
Then
\begin{equation}
K_i(-r)=\operatorname{Cub}_i\bigl(B_i(r)\bigr).
\label{eq:inneroffsetidentity}
\end{equation}
Thus, as the threshold increases from negative values toward zero, the filtered bone grows outward from voxel centers deepest in the bone phase until $K_i(0)=\operatorname{Cub}_i(B_i)$. Positive thresholds subsequently admit marrow voxels lying increasingly farther from bone.

At each threshold $t$, define
\begin{equation}
\beta_{k,i}(t)=\dim_{\mathbb F_2}H_k\bigl(K_i(t);\mathbb F_2\bigr).
\label{eq:betti}
\end{equation}
Whenever $u\le t$, the inclusion $K_i(u)\hookrightarrow K_i(t)$ induces a linear map
\[
\rho^{(k)}_{u,t}:H_k\bigl(K_i(u);\mathbb F_2\bigr)
\longrightarrow H_k\bigl(K_i(t);\mathbb F_2\bigr).
\]
The homology groups together with these inclusion-induced maps form the $k$th persistence module. Because the filtration is finite, this module decomposes into interval summands \cite{Zomorodian2004,Kaczynski2004}. A finite interval $[b,d)$ records a homology class born at $b$ and dying at $d>b$, with persistence $\ell=d-b$.

For vectorization, the finite intervals in dimension $k$ are represented in birth and persistence coordinates by the multiset
\[
D_{k,i}=\{(b,\ell):\ell=d-b>0\}.
\]
Intervals of zero persistence are omitted. The terminal ROI complex is a connected digital cylinder, so its unique essential $H_0$ class is excluded from the finite representation used for prediction. Here, \emph{essential} means that the class survives every subsequent finite threshold.

The signed distance construction gives finite $H_0$ intervals a particularly direct geometric interpretation. If a class is born at $b<0$, Equation~\eqref{eq:inneroffsetidentity} implies that its component first appears among bone voxels at least $-b$ micrometers from marrow voxel centers. If its death satisfies $d\le0$, its merger occurs entirely within the nested interior bone sets. If $d>0$, the component survives through the original bone boundary and merges only after marrow voxels enter the filtration. $H_0$ persistence therefore records both the depth at which connected interior regions appear and the scales at which those regions become connected.

This information is related to, but is not equivalent to, trabecular thickness. Local thickness is defined through maximal spheres contained in bone \cite{Hildebrand1997}, whereas $H_0$ persistence records connectivity changes across an entire nested family of interior subcomplexes. Likewise, $H_1$ and $H_2$ describe cycles of the evolving filtered complex, not intrinsically less relevant features. Their association with strength is therefore an empirical question rather than something determined by homological dimension alone.

\subsection{Persistence images}
\label{sec:persistence_images}

Persistence diagrams contain different numbers of intervals across specimens and therefore do not directly provide predictors of fixed dimension. Each diagram $D_{k,i}$ is mapped to a persistence image by replacing every point in birth and persistence coordinates with a weighted Gaussian and integrating the resulting surface over a fixed rectangular grid \cite{JMLR:v18:16-337}.

The isotropic Gaussian density with bandwidth $\sigma>0$ is
\[
\phi_\sigma(u,v)=\frac{1}{2\pi\sigma^2}
\exp\!\left(-\frac{u^2+v^2}{2\sigma^2}\right).
\]
For homology dimension $k$, let $\omega_k(\ell)=\min\{\ell/\tau_k,1\}$, where $\tau_k>0$ is a fixed persistence scale. For a rectangular grid cell $P_{rs}=[u_r,u_{r+1}]\times[v_s,v_{s+1}]$, define
\begin{equation}
p^{(k)}_{i,rs}=
\sum_{(b,\ell)\in D_{k,i}}\omega_k(\ell)
\iint_{P_{rs}}\phi_\sigma(u-b,v-\ell)\,\mathrm du\,\mathrm dv.
\label{eq:pi}
\end{equation}
Each coordinate therefore records integrated Gaussian mass rather than a sampled surface height. Since $D_{k,i}$ is finite, linearity of integration permits the finite sum and integral to be interchanged.

Let $\Phi$ denote the standard normal cumulative distribution function. Gaussian separability gives the exact contribution of one persistence point to one cell:
\begin{align}
\iint_{P_{rs}}\phi_\sigma(u-b,v-\ell)\,\mathrm du\,\mathrm dv
={}&\left[\Phi\!\left(\frac{u_{r+1}-b}{\sigma}\right)
-\Phi\!\left(\frac{u_r-b}{\sigma}\right)\right]\nonumber\\
&\times\left[\Phi\!\left(\frac{v_{s+1}-\ell}{\sigma}\right)
-\Phi\!\left(\frac{v_s-\ell}{\sigma}\right)\right].
\label{eq:cellintegral}
\end{align}
Gaussian mass outside the finite grid domain is not represented.

The primary analysis used $\sigma=25~\mu\mathrm m$. The grid resolutions and persistence weight saturation scales for each homology dimension are given in Table~\ref{tab:pi_grids}. These choices were fixed from pooled baseline diagrams before strength outcomes were accessed and were then held fixed. Complete numerical conventions are reported in \ref{app:images}.

\begin{table}[htbp]
\centering
\caption{Persistence image discretization by homology dimension. The number of coordinates is the product of the numbers of birth and persistence cells.}
\label{tab:pi_grids}
\begin{tabular}{lrrr}
\toprule
Dimension & Birth $\times$ persistence cells & Coordinates & $\tau_k$ ($\mu\mathrm m$)\\
\midrule
$H_0$ & $30\times20$ & 600 & 374.703\\
$H_1$ & $40\times20$ & 800 & 715.692\\
$H_2$ & $44\times20$ & 880 & 520.009\\
\bottomrule
\end{tabular}
\end{table}

Let $p_i^{(k)}$ be the vector obtained by arranging all coordinates $p^{(k)}_{i,rs}$ in a fixed order. Writing $\oplus$ for vector concatenation, the original primary topological representation is
\begin{equation}
\mathcal P(B_i;R_i)=p_i=p_i^{(0)}\oplus p_i^{(1)}\oplus p_i^{(2)}\in\R^{2280}.
\label{eq:combinedpi}
\end{equation}
The three homology dimensions were also evaluated separately as exploratory representations. The pooled grid construction used the complete set of baseline images but no strength outcomes; fold-specific variance filtering, centering, and scaling were still recomputed within each training set.

Figure~\ref{fig:pi_schematic} illustrates the mapping from a finite persistence diagram to a persistence image.

\begin{figure}[tbp]
\centering
\includegraphics[width=0.82\linewidth]{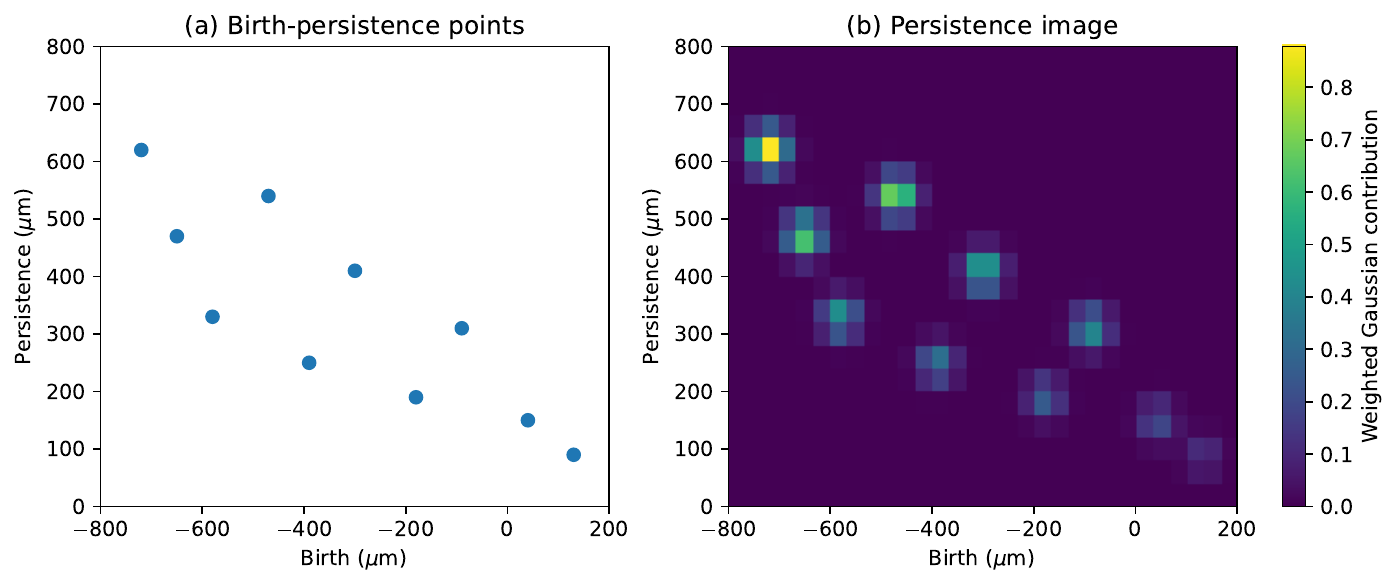}
\caption{Schematic construction of a persistence image. Each point contributes a Gaussian surface weighted by persistence, and integration over each rectangular grid cell produces one coordinate of the resulting vector. The displayed points are illustrative rather than observations from the study cohort.}
\label{fig:pi_schematic}
\end{figure}

\subsection{Directional multiparameter filtrations}
\label{sec:directional_multifiltrations}

The signed distance construction assigns one scalar geometric scale to each voxel. To encode directional geometry, each bone voxel is instead assigned a vector in $\mathbb R^3$, using the coordinatewise order introduced in Section~\ref{sec:cubical_homology}. The result is a three-parameter filtration rather than three unrelated scalar filtrations.

The construction uses two local geometric ingredients: bone occupancy inside a cone aligned with a selected physical direction and the covariance structure of nearby bone. We first state the reference construction reproduced from \cite{Topolnicki2026Directional}, because its dependence on a transverse coordinate choice motivates the invariant reformulation.

\subsubsection{Directional cone coordinate}

Fix a specimen $i$ and a unit direction
\[
d\in\mathbb S^2=\{u\in\mathbb R^3:\|u\|_2=1\}.
\]
Let $R_c=75~\mu\mathrm m$ be the cone radius and $H=150~\mu\mathrm m$ its axial half extent. For voxel indices $x,y\in R_i$, define the physical displacement
\[
r_{xy}=h_i(y-x)\in\mathbb R^3.
\]
With $I$ the $3\times3$ identity, the symmetric double cone, restricted to the ROI, centered at $x$ and oriented along $d$ is
\begin{equation}
\Gamma_{i,d}(x)=
\left\{y\in R_i:
|r_{xy}^{\mathsf T}d|\le H
\text{ and }
\|(I-dd^{\mathsf T})r_{xy}\|_2
\le\frac{R_c}{H}|r_{xy}^{\mathsf T}d|
\right\}.
\label{eq:cone_support}
\end{equation}
The projection $dd^{\mathsf T}r_{xy}$ is the component parallel to $d$, while $(I-dd^{\mathsf T})r_{xy}$ is the component in the perpendicular plane $d^\perp$. Hence the first inequality bounds axial displacement and the second makes the transverse radius grow linearly with axial distance.

For a bone voxel $x\in B_i$, define the \emph{directional cone deficit}
\begin{equation}
C_{i,d}(x)=1-\frac{|\Gamma_{i,d}(x)\cap B_i|}{|\Gamma_{i,d}(x)|}.
\label{eq:cone_coordinate}
\end{equation}
Then $0\le C_{i,d}(x)\le1$, and smaller values correspond to greater bone occupancy inside the directional cone. The quantity depends on the physical direction $d$ but not on a choice of basis for $d^\perp$. The cohort analysis uses $d\in\{e_x,e_y,e_z\}$, with $e_z$ corresponding to the loading direction.

\subsubsection{Reference coordinates and frame dependence}
\label{sec:reference_directional}

Local orientation is estimated from the spatial distribution of neighboring bone voxels. Let $R_p=75~\mu\mathrm m$. The hard neighborhood used in the original reproduction is
\begin{equation}
w_x^{\mathrm H}(y)=\mathbf 1\{\|r_{xy}\|_2\le R_p\},
\label{eq:hard_weight}
\end{equation}
whereas the later smooth weighting uses $\sigma_p=R_p/\sqrt5$ and is
\begin{equation}
w_x^{\mathrm G}(y)=
\exp\!\left(-\frac{\|r_{xy}\|_2^2}{2\sigma_p^2}\right)
\mathbf 1\{\|r_{xy}\|_2\le4\sigma_p\}.
\label{eq:gaussian_weight}
\end{equation}
Before truncation, this choice of $\sigma_p$ matches the radial second moment of a uniform solid ball of radius $R_p$; the Gaussian weight at the truncation radius is $\exp(-8)$.

For either radial weighting, set
\[
W_x=\sum_{y\in B_i}w_x(y).
\]
Then set
\[
\mu_x=\frac{1}{W_x}\sum_{y\in B_i}w_x(y)r_{xy}.
\]
The weighted covariance is
\begin{equation}
\Sigma_x=
\frac{1}{W_x}\sum_{y\in B_i}w_x(y)
(r_{xy}-\mu_x)(r_{xy}-\mu_x)^{\mathsf T}.
\label{eq:local_covariance}
\end{equation}
The matrix $\Sigma_x$ is symmetric positive semidefinite. Let
\[
\lambda_1(x)\ge\lambda_2(x)\ge\lambda_3(x)\ge0
\]
be its eigenvalues. Where the leading eigenspace is one-dimensional, choose a unit eigenvector $v_1(x)$ for $\lambda_1(x)$ and define
\begin{equation}
P_q(x)=1-|v_1(x)^{\mathsf T}e_q|.
\label{eq:reference_orientation}
\end{equation}
The absolute value removes the sign ambiguity $v_1\leftrightarrow-v_1$, but it does not resolve the lack of uniqueness when the largest eigenvalue is repeated.

Writing $C_q(x)=C_{i,e_q}(x)$, the reproduced directional coordinates can be written as the single map
\begin{equation}
F_q^{\mathrm{leg}}(x)=
\begin{cases}
(C_x(x),P_z(x),P_x(x)), & q=x,\\
(C_y(x),P_z(x),P_y(x)), & q=y,\\
(C_z(x),P_y(x),P_z(x)), & q=z.
\end{cases}
\label{eq:legacy_directional_filters}
\end{equation}
The first coordinate is directional cone occupancy and the third measures alignment with the requested axis. The second coordinate, however, retains one selected component in the plane perpendicular to that axis. Once $d$ is fixed, the physical plane $d^\perp$ admits infinitely many orthonormal bases. Rotating that transverse basis can therefore change the second coordinate while leaving $d$ and the physical bone neighborhood unchanged. Replacing Equation~\eqref{eq:hard_weight} by the smoother Equation~\eqref{eq:gaussian_weight} changes the covariance estimate but does not remove this dependence on the transverse frame.

\subsection{Frame-invariant directional coordinates}
\label{sec:frame_invariant_directional}

The invariant construction replaces selected components of a principal eigenvector by scalars determined by the covariance tensor itself and the selected physical direction. In this subsection, $\Sigma_x$ denotes the Gaussian weighted covariance obtained from Equation~\eqref{eq:local_covariance} using $w_x^{\mathrm G}$.

\begin{definition}[Invariant local coordinates]
\label{def:invariant_coordinates}
For a bone voxel $x$ with $\operatorname{tr}\Sigma_x>0$ and a unit direction $d$, define the \emph{directional covariance deficit}
\begin{equation}
A_{i,d}(x)=1-
\sqrt{\frac{d^{\mathsf T}\Sigma_xd}{\operatorname{tr}\Sigma_x}},
\label{eq:invariant_coordinates}
\end{equation}
and the \emph{principal axis degeneracy}
\begin{equation}
S_i(x)=\frac{\lambda_2(x)}{\lambda_1(x)}.
\label{eq:axis_degeneracy}
\end{equation}
If the weighted neighborhood contains fewer than two bone voxels with positive weight, or if $\operatorname{tr}\Sigma_x=0$, set $A_{i,d}(x)=S_i(x)=1$.
\end{definition}

Because $\Sigma_x$ is positive semidefinite,
\[
0\le d^{\mathsf T}\Sigma_xd\le\operatorname{tr}\Sigma_x,
\]
so $0\le A_{i,d}(x)\le1$. A small $A_{i,d}$ means that a large fraction of local covariance lies along $d$. Likewise, $0\le S_i(x)\le1$. The value $S_i=0$ corresponds to a rank one covariance, whereas $S_i\to1$ indicates that the two largest eigenvalues become comparable and the leading local axis becomes less distinct.

The next proposition connects the invariant alignment coordinate to the principal axis coordinate in the idealized rank one case.

\begin{proposition}
\label{prop:rankone}
Suppose $\Sigma_x=\lambda vv^{\mathsf T}$ with $\lambda>0$ and $\|v\|_2=1$. Then $A_{i,d}(x)=1-|v^{\mathsf T}d|$ and $S_i(x)=0$.
\end{proposition}

\begin{proof}
Since $\operatorname{tr}(\lambda vv^{\mathsf T})=\lambda$ and $d^{\mathsf T}\Sigma_xd=\lambda(v^{\mathsf T}d)^2$, Equation~\eqref{eq:invariant_coordinates} gives
\[
A_{i,d}(x)=1-\sqrt{(v^{\mathsf T}d)^2}=1-|v^{\mathsf T}d|.
\]
The remaining eigenvalues of the rank one matrix are zero, so $S_i(x)=0$.
\end{proof}

Thus the frame-invariant descriptor assigned to a voxel is
\begin{equation}
F_{i,d}^{\mathrm{inv}}(x) = \bigl(C_{i,d}(x),A_{i,d}(x),S_i(x)\bigr).
\label{eq:invariant_filter}
\end{equation}

The relevant notion of frame invariance is a passive change of orthonormal coordinates that keeps the physical direction $d$ fixed. Let
\[
O(3)=\{Q\in\mathbb R^{3\times3}:Q^{\mathsf T}Q=I\}.
\]

\begin{proposition}[Transverse-frame invariance]
\label{prop:frameinvariance}
Let $Q\in O(3)$ satisfy $Qd=d$. Express the same physical neighborhood in the transformed orthonormal frame by
$r'_{xy}=Qr_{xy}$ and $\Sigma'_x=Q\Sigma_xQ^{\mathsf T}$. Then $C'_{i,d}(x)=C_{i,d}(x)$, $A'_{i,d}(x)=A_{i,d}(x)$, and $S'_i(x)=S_i(x)$. Consequently, $F_{i,d}^{\mathrm{inv}}(x)$ is independent of the orthonormal basis chosen in $d^\perp$.
\end{proposition}

\begin{proof}
Let $P_d=I-dd^{\mathsf T}$ be the orthogonal projector onto $d^\perp$. Since $Q$ is orthogonal and $Qd=d$, also $Q^{\mathsf T}d=d$ and $QP_d=P_dQ$. Hence
\[
(r'_{xy})^{\mathsf T}d=r_{xy}^{\mathsf T}d
\]
and
\[
\|P_dr'_{xy}\|_2
=\|P_dQr_{xy}\|_2
=\|QP_dr_{xy}\|_2
=\|P_dr_{xy}\|_2.
\]
The two inequalities defining the cone in Equation~\eqref{eq:cone_support} are therefore unchanged, so $C'_{i,d}(x)=C_{i,d}(x)$.

Moreover, $\Sigma'_x=Q\Sigma_xQ^{\mathsf T}$ is orthogonally similar to $\Sigma_x$, so the two matrices have the same eigenvalues and the same trace. Finally,
\[
d^{\mathsf T}\Sigma'_xd
=d^{\mathsf T}Q\Sigma_xQ^{\mathsf T}d
=d^{\mathsf T}\Sigma_xd.
\]
Equations~\eqref{eq:invariant_coordinates} and \eqref{eq:axis_degeneracy} then give the remaining two identities.
\end{proof}

Proposition~\ref{prop:frameinvariance} concerns two coordinate descriptions of the same physical neighborhood and is therefore exact. It is different from physically rotating a voxelized image and resampling it on a Cartesian lattice; the latter changes the discrete sampled geometry through interpolation and voxelization and is not expected to be exactly invariant.

\subsection{Cubical extension of the directional coordinates}
\label{sec:directional_cubical_extension}

The vector $F_{i,d}^{\mathrm{inv}}(x)$ is initially attached only to a bone voxel, equivalently to its top-dimensional cube $Q_i(x)$. To obtain a filtration by cubical \emph{subcomplexes}, values must also be assigned to lower-dimensional faces in a way that guarantees closure under taking faces.

Let $\gamma=1.25$ and assign to each top-dimensional ROI cube
\begin{equation}
f_{i,d}(Q_i(x))=
\begin{cases}
F_{i,d}^{\mathrm{inv}}(x), & x\in B_i,\\
(\gamma,\gamma,\gamma), & x\in R_i\setminus B_i.
\end{cases}
\label{eq:topcell_filter}
\end{equation}
The value $\gamma>1$ is a terminal numerical level, not a biological parameter. Since all bone coordinates lie in $[0,1]$, marrow enters only after every level assigned to bone. Cubes outside $R_i$ are absent from the ambient complex. The legacy construction uses the same extension after replacing $F_{i,d}^{\mathrm{inv}}$ by the corresponding triple in Equation~\eqref{eq:legacy_directional_filters}.

For a cubical cell $\sigma\in K_i^{\mathrm{ROI}}$, let
\[
\mathcal Q_i(\sigma)=
\{Q_i(x):x\in R_i\text{ and }\sigma\text{ is a face of }Q_i(x)\}
\]
be the set of incident top-dimensional ROI cubes. Extend coordinate $j\in\{1,2,3\}$ to every cell by the face minimum
\begin{equation}
\bar f_{i,d,j}(\sigma)
=
\min_{Q\in\mathcal Q_i(\sigma)}f_{i,d,j}(Q).
\label{eq:coordinatewise_birth}
\end{equation}
For $\mathbf t=(t_1,t_2,t_3)\in\mathbb R^3$, define
\begin{equation}
K_{i,d}(\mathbf t)
=
\{\sigma\in K_i^{\mathrm{ROI}}:
\bar f_{i,d,j}(\sigma)\le t_j\text{ for }j=1,2,3\}.
\label{eq:directional_complex}
\end{equation}

\begin{proposition}
\label{prop:subcomplex}
For every $\mathbf t\in\mathbb R^3$, $K_{i,d}(\mathbf t)$ is a cubical subcomplex of $K_i^{\mathrm{ROI}}$. If $\mathbf s\preceq\mathbf t$, then
\[
K_{i,d}(\mathbf s)\subseteq K_{i,d}(\mathbf t).
\]
Hence $(K_{i,d}(\mathbf t))_{\mathbf t\in\mathbb R^3}$ is a three-parameter cubical filtration.
\end{proposition}

\begin{proof}
Let $\sigma\in K_{i,d}(\mathbf t)$ and let $\tau$ be a face of $\sigma$. Every top-dimensional cube containing $\sigma$ also contains $\tau$, so
\[
\mathcal Q_i(\sigma)\subseteq\mathcal Q_i(\tau).
\]
Taking a minimum over the larger set can only decrease the value, giving
\[
\bar f_{i,d,j}(\tau)\le\bar f_{i,d,j}(\sigma)\le t_j
\]
for every $j$. Thus $\tau\in K_{i,d}(\mathbf t)$, proving closure under faces. The monotonicity statement follows immediately from $s_j\le t_j$ for every coordinate.
\end{proof}

An important consequence of Equation~\eqref{eq:coordinatewise_birth} is that the three minima for a lower-dimensional cell need not be attained by the same incident voxel cube. A cell may therefore satisfy all three threshold inequalities even when no single adjacent voxel satisfies them jointly. Accordingly, $K_{i,d}(\mathbf t)$ is the cubical subcomplex defined by the coordinatewise face minimum extension; in general it is not merely the closure of a jointly thresholded voxel mask.

\subsection{Sampled Betti tensors}
\label{sec:sampled_betti_tensors}

The complete three-parameter persistence module is not computed. Instead, the filtration is sampled on the finite poset
\[
\mathcal G=\{0,\ldots,8\}^3
\]
with coordinatewise order. Thus the resulting arrays are samples of the multivariate Betti function, not a complete multiparameter persistence invariant.

For fixed specimen $i$, direction $d$, and filtration coordinate $j$, let $m_j=\min_Q f_{i,d,j}(Q)$ and $M_j=\max_Q f_{i,d,j}(Q)$, where the extrema are taken over top-dimensional ROI cubes. Set
\begin{equation}
\delta_j=\frac{M_j-m_j}{8}.
\label{eq:grid_width}
\end{equation}
Because marrow has terminal value $\gamma=1.25$ and bone coordinates lie in $[0,1]$, $M_j>m_j$ and hence $\delta_j>0$.

Assign each top-dimensional cube the discrete coordinate
\begin{equation}
b_j(Q)=
\left\lfloor\frac{f_{i,d,j}(Q)-m_j}{\delta_j}\right\rfloor
\in\{0,\ldots,8\}.
\label{eq:top_bin}
\end{equation}
For $a<8$, the condition $b_j(Q)\le a$ is equivalent to
\[
f_{i,d,j}(Q)<m_j+(a+1)\delta_j,
\]
whereas index $8$ includes every top-dimensional cube. Thus an index denotes inclusion of all bins up to that index; it should not be interpreted as evaluation at the lower bin edge $m_j+a\delta_j$.

For a lower-dimensional cell $\sigma$, set
\begin{equation}
b_j(\sigma)=\min_{Q\in\mathcal Q_i(\sigma)}b_j(Q).
\label{eq:cell_bin}
\end{equation}
Since the floor function is nondecreasing and the incident set is finite, quantization commutes with the coordinatewise minimum:
\[
\min_{Q\in\mathcal Q_i(\sigma)}
\left\lfloor\frac{f_{i,d,j}(Q)-m_j}{\delta_j}\right\rfloor
=
\left\lfloor
\frac{\min_{Q\in\mathcal Q_i(\sigma)}f_{i,d,j}(Q)-m_j}{\delta_j}
\right\rfloor.
\]

For $(a,b,c)\in\mathcal G$, define the sampled cubical complex
\begin{equation}
K_{i,d}[a,b,c]
=
\{\sigma:b_1(\sigma)\le a,\ b_2(\sigma)\le b,\ b_3(\sigma)\le c\}.
\label{eq:sampled_complex}
\end{equation}
The $9^3=729$ sampled complexes inherit the coordinatewise nesting of the underlying three-parameter filtration.

For homology dimension $q\in\{0,1,2\}$, the \emph{directional Betti tensor} is
\begin{equation}
\mathcal B_{q,i}^{d}(a,b,c)
=
\dim_{\mathbb F_2}H_q\bigl(K_{i,d}[a,b,c];\mathbb F_2\bigr).
\label{eq:betti_tensor}
\end{equation}
Each tensor $\mathcal B_{q,i}^{d}\in\mathbb N_0^{9\times9\times9}$ contains 729 Betti numbers. It records the dimension of homology at each sampled parameter value, but not the inclusion-induced maps between different parameter values. Therefore it does not by itself determine which particular classes persist, merge, or disappear across the parameter grid. In addition, because $(m_j,M_j)$ are estimated separately for each specimen and direction, a common tensor index represents a common \emph{relative} filtration level rather than a common numerical threshold.

For the three cardinal directions, define the transverse mean by
\begin{equation}
\mathcal B_{q,i}^{\perp}
=
\frac12\bigl(\mathcal B_{q,i}^{x}+\mathcal B_{q,i}^{y}\bigr).
\label{eq:transverse_mean}
\end{equation}
The axial contrast is
\begin{equation}
\Delta\mathcal B_{q,i}^{\mathrm{ax}}
=
\mathcal B_{q,i}^{z}-\mathcal B_{q,i}^{\perp}.
\label{eq:axial_contrast}
\end{equation}
These are derived tensor summaries and need not themselves be Betti functions of cubical complexes; in particular, $\mathcal B_{q,i}^{\perp}$ may have noninteger entries. When all three directional tensors are retained as predictors, no averaging is performed: the vectorized block is
\[
\operatorname{vec}\mathcal B_{q,i}^{x}
\oplus
\operatorname{vec}\mathcal B_{q,i}^{y}
\oplus
\operatorname{vec}\mathcal B_{q,i}^{z},
\]
with $3\times729=2187$ coordinates for each homology dimension.

The implementation computes $\mathcal B_1$ through the Euler--Poincar\'e identity. For a finite cubical complex $K$, let $n_k(K)$ be the number of its $k$-cells. Then
\[
\chi(K)=n_0(K)-n_1(K)+n_2(K)-n_3(K)
\]
and
\[
\chi(K)=\sum_{k=0}^{3}(-1)^k\dim_{\mathbb F_2}H_k(K;\mathbb F_2).
\]
For the finite cubical subcomplexes considered here, $H_3=0$. Indeed, any nonempty finite collection of three-dimensional voxel cubes in $\mathbb R^3$ has at least one exposed square face, so a nonzero $3$-chain cannot have zero cubical boundary. Hence $\ker\partial_3=0$.

Define the sampled Euler tensor
\[
\mathcal E_i^d(a,b,c)=\chi\bigl(K_{i,d}[a,b,c]\bigr).
\]
Then
\begin{equation}
\mathcal E_i^d
=
\mathcal B_{0,i}^d-\mathcal B_{1,i}^d+\mathcal B_{2,i}^d,
\label{eq:euler_betti_tensor}
\end{equation}
or equivalently
\begin{equation}
\mathcal B_{1,i}^d
=
\mathcal B_{0,i}^d+\mathcal B_{2,i}^d-\mathcal E_i^d.
\label{eq:b1_from_euler}
\end{equation}
The implementation computes $\mathcal B_0$ from connectivity in the filtered 1-skeleton using union--find, $\mathcal B_2$ from connectivity of the dual complement after identifying the exterior component, and $\mathcal E$ independently from cubical cell counts. Equation~\eqref{eq:b1_from_euler} then gives $\mathcal B_1$.

Finally, Proposition~\ref{prop:frameinvariance} propagates through the complete sampled construction. A transverse change of orthonormal coordinates that fixes $d$ leaves the values on top-dimensional cells unchanged; therefore the coordinatewise face minima, extrema for each specimen, bins, sampled subcomplexes, and their Betti and Euler tensors are all unchanged. The sampled representation is thus exactly transverse-frame invariant for a fixed voxel geometry, while remaining subject to ordinary resampling error under physical rotation of the voxelized image.
\end{revision}

\section{Methods}
\label{sec:methods}

\subsection{Image construction}
\label{sec:image_construction}

Of 59 available microCT volumes, 24 core identifiers could be matched to recorded apparent compressive strength. These comprised 19 fracture (FX), three nonfracture (NF), and two osteoarthritis (OA) cores. For specimen $i$, the prediction target was
\[
y_i=F_{\max,i}/A_{0,i},
\]
where $F_{\max,i}$ is the maximum recorded compressive force and $A_{0,i}$ is the recorded area of the cross section. Specimen collection, consent, and mechanical testing are described in the source study and thesis \cite{Jin2017,Jin_2016}. The analysis therefore concerns measured strength of excised cores; clinical fracture status was not the prediction target.

Scans were assigned random identifiers QC001--QC024 before image analysis. Specimen identities, clinical labels, and mechanical outcomes were withheld until the primary image representations had been fixed. The image $Z$ axis corresponded to the loading direction. A central cylinder of radius $2.75~\mathrm{mm}$ and nominal height $8.50~\mathrm{mm}$ defined the ROI. Its axial position was determined from slice intensity variation, with fixed trimming of the specimen ends. Voxel spacing was effectively isotropic within each scan and was retained in physical distance calculations.

\begin{figure}[htbp]
\centering
\includegraphics[width=0.88\linewidth]{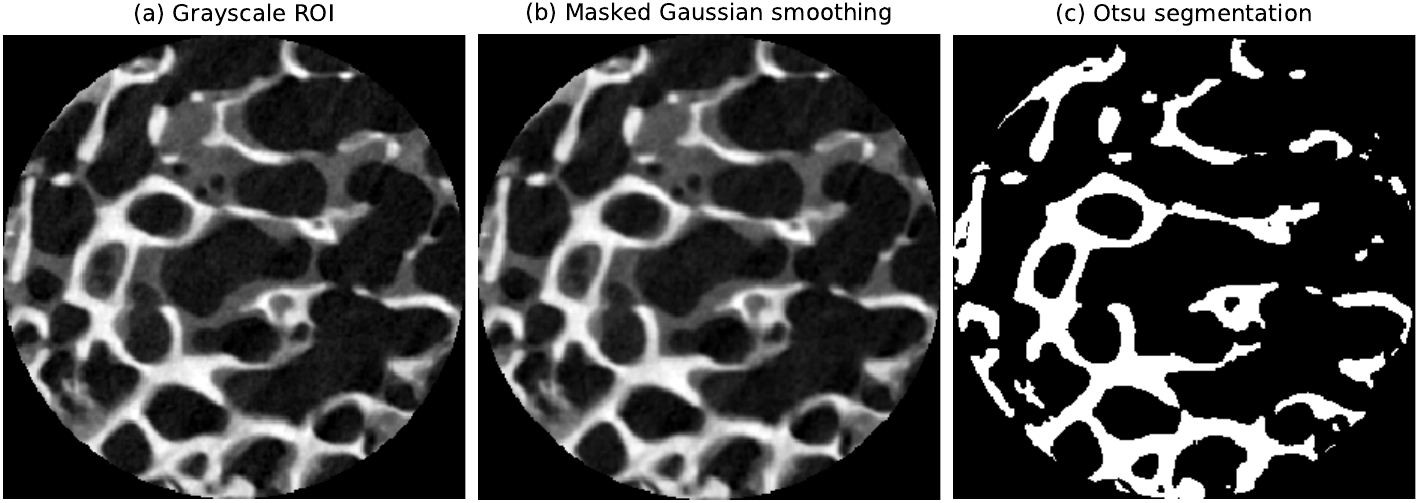}
\caption{Image construction for a representative axial slice: grayscale image, normalized masked Gaussian smoothing, and binary segmentation by a single three-dimensional Otsu threshold. The cylindrical ROI defines the analysis domain.}
\label{fig:segmentation}
\end{figure}

For grayscale image $I_i$ and ROI indicator $M_i$, normalized masked smoothing for $x\in R_i$ was
\[
I_{\sigma_I,i}(x)=
\frac{[G_{\sigma_I}*(I_iM_i)](x)}{[G_{\sigma_I}*M_i](x)}.
\]
The Gaussian standard deviation was $\sigma_I=0.50$ voxel, with a kernel radius of two voxels. A single three-dimensional Otsu threshold $T_i$, estimated from a 256-bin histogram inside the ROI, yields the binary image whose bone voxels are given by 
\[
B_i=\{x\in R_i:I_{\sigma_I,i}(x)>T_i\}.
\]
No component filtering, hole filling, opening, closing, or manual topological correction was applied. Figure~\ref{fig:segmentation} illustrates this construction. Appendix~\ref{app:images} specifies the axial detection rule and morphometric boundary conventions, including the inscribed prism used for anisotropy estimation.

\begin{revision}
\subsection{Study design and validation groups}
\label{sec:analysis_hierarchy}
The analyses were developed at different stages and should not be read as a single prospectively specified experiment. Table~\ref{tab:analysis_hierarchy} summarizes their sequence and purpose. The original benchmark compared the combined persistence image representation with conventional morphometry. The separate homology dimensions were exploratory and motivated the later emphasis on $H_0$. The directional reproduction, the check of frame dependence, the invariant reformulation, the structural comparisons, and the mechanical experiments were developed later using the same cohort. Fixing each later representation before linking it to outcomes limited further tuning, but did not make those analyses prospective.

\begin{table}[htbp]
\centering
\color{revisioncolor}
\caption{Sequence and purpose of the analyses. The first row is the original primary benchmark; the remaining rows are exploratory or later analyses of the same cohort.}
\label{tab:analysis_hierarchy}
\begin{tabularx}{\linewidth}{L{0.24\linewidth}L{0.37\linewidth}Y}
\toprule
Analysis & Representation or comparison & Purpose\\
\midrule
Original benchmark & Morphometry; combined $H_0\oplus H_1\oplus H_2$ persistence images; concatenation & Strength prediction\\
Homology dimensions & Separate signed distance $H_0$, $H_1$, and $H_2$ persistence images & Compare dimensions\\
Directional verification & Reference Betti tensors; $90^\circ$ transverse roll & Test dependence on the transverse frame\\
Invariant construction & Frame-invariant $\mathcal B_q^d$ tensors & Remove transverse frame dependence\\
Structural comparisons & Skeleton, graph, organization, orientation, and surface tensor descriptors & Compare geometric representations\\
Mechanical experiments & Equal removal and a pair matched in morphometry & Test mechanical interpretations of signed distance topology\\
\bottomrule
\end{tabularx}
\end{table}

Bandwidth changes and addition of the principal fabric angle remained specified sensitivity analyses of the original benchmark. Later paired prediction contrasts are taken from a common run whenever possible so that each difference compares models evaluated under the same procedure.
\end{revision}

The available partition contained $G=13$ validation groups: 11 pairs of cores and two single cores. These identifiers have not been verified as donor identities. Each group was kept intact in every validation split and bootstrap sample. If $g(i)$ is the group containing specimen $i$, and $n_g$ is its size, the analysis weight was $w_i=1/n_{g(i)}$. Consequently, each group had total weight one and $\sum_iw_i=G$. Groupwise validation limits overlap across the assigned groups; it does not establish independence between donors whose grouping is unknown.

\subsection{Nested prediction and uncertainty}

\subsubsection{Preprocessing in training folds}

Variance filtering, centering, and scaling were estimated anew in every inner and outer training set. For a predictor block on training set $T$, let $s_{ij}$ denote the support values. Their weighted mean is
\[
\bar s_j=\frac{\sum_{i\in T}w_is_{ij}}{\sum_{i\in T}w_i}.
\]
The corresponding weighted variance is
\[
v_j=\frac{\sum_{i\in T}w_i(s_{ij}-\bar s_j)^2}{\sum_{i\in T}w_i}.
\]
The active set was $J_T=\{j:v_j>10^{-12}\max_kv_k\}$. Persistence image support used the image coordinates themselves. \textcolor{revisioncolor}{For the directional predictors normalized by ROI size, support values were determined from the corresponding raw count tensors before normalization.}

For the predictor values $x_{ij}$ entering an active block, let $\bar x_j$ be the weighted training mean. One scale was assigned to the entire block:
\[
\rho_T^2= \frac{\sum_{i\in T}w_i\sum_{j\in J_T}(x_{ij}-\bar x_j)^2}{(\sum_{i\in T}w_i)|J_T|}.
\]
The transformed predictor was then
\[
\widetilde x_{ij}=\frac{x_{ij}-\bar x_j}{\rho_T}.
\]
Topological blocks were processed separately before concatenation. Scalar morphometric predictors were standardized individually and their block was divided by the square root of its dimension. Each transformation was applied unchanged to the corresponding validation observations. The persistence image grid itself had been constructed from pooled baseline images before outcome access, as specified in Section~\ref{sec:framework}. It was not reconstructed within these folds.

\subsubsection{Model fitting and evaluation}

For transformed predictors $x_i$, weighted ridge regression \cite{HoerlKennard1970} solved
\[
(\widehat\beta_0,\widehat\beta)= \operatorname*{arg\,min}_{\beta_0,\beta}\left\{\sum_{i\in T}w_i(y_i-\beta_0-x_i^{\mathsf T}\beta)^2 +\alpha\|\beta\|_2^2\right\}.
\]
The intercept was unpenalized. The outer loop withheld one complete group. Within the remaining groups, inner leave-one-group-out validation selected $\alpha$ from the 49 values $10^{-6+j/4}$, $j=0,\ldots,48$, by mean squared error with equal weight per validation group. The selected model was refitted on the complete outer training set and applied to the held out group. Training mean prediction and ordinary linear regression on BV/TV alone provided simpler references.

Let $\widehat y_i^{\,x}$ be the outer prediction from representation $x$. The principal error measure was
\begin{equation}
\mathcal R_G(x)=
\left\{\frac{1}{G}\sum_{i=1}^{24}w_i
(y_i-\widehat y_i^{\,x})^2\right\}^{1/2}.
\label{eq:risk}
\end{equation}
The original primary contrasts were $\Delta_{\mathrm{rep}} =\mathcal R_G(p)-\mathcal R_G(m)$ and $\Delta_{\mathrm{add}} =\mathcal R_G(m\oplus p)-\mathcal R_G(m)$. Negative values favor the candidate representation or its addition. Mean absolute error and $Q_G^2$ are defined in Appendix~\ref{app:prediction}.

Uncertainty intervals used 10,000 bootstrap resamples of the 13 groups, with the same resampled groups for both models in a paired comparison. Metrics were recalculated from fixed outer predictions; models were not refitted during the bootstrap. The 2.5th and 97.5th percentiles therefore describe variation in the observed prediction errors conditional on the fitted procedure. They do not include uncertainty from repeating representation development or selecting a homology dimension. Later exploratory comparisons were not adjusted for multiplicity. All reported differences retain the precision of the original analyses, so subtraction of rounded table entries can differ in the last decimal.

\subsection{Representation similarity}

The later families of descriptors summarized the thinned skeleton (eight variables), axial routes and graph transport (five), a plate and rod organization proxy (one), local orientation variation (three), and the shape of the surface-normal Minkowski tensor $W_1^{0,2}$ (two). Their definitions and complete comparisons appear in Appendix~\ref{app:comparators}. The organization proxy uses local geometric samples rather than individual trabeculae reconstructed by ITS. The graph transport model uses branch lengths rather than tissue mechanics. These distinctions delimit the representations being tested.

Representation agreement was assessed separately from prediction. For column-centered specimen matrices $X_c$ and $Z_c$, linear centered-kernel alignment was \cite{Kornblith2019CKA}
\begin{equation}
\operatorname{CKA}(X,Z)= \frac{\|X_c^{\mathsf T}Z_c\|_F^2}{\|X_c^{\mathsf T}X_c\|_F\,\|Z_c^{\mathsf T}Z_c\|_F}.
\label{eq:cka}
\end{equation}
whenever the denominator is positive. Here, $\|\cdot\|_F$ is the Frobenius norm. We also computed Spearman correlation between the upper triangles of the two Euclidean interspecimen distance matrices. Both summaries describe the geometry induced on this cohort. Neither tests incremental prediction.

\begin{revision}
\subsection{Haufe activation patterns}
\label{sec:b0_haufe_methods}
The same linear model interpretation is used in two different feature spaces later in the manuscript, so the transformation is stated once here. For an outer training fold $T$, let $X_T$ be a centered predictor block on its original feature scale and let $s_T$ be the centered fitted scalar score from the corresponding ridge model. On an active feature coordinate $j$, define the weighted \emph{Haufe activation pattern} \cite{Haufe2014} by
\begin{equation}
\mathcal H_T(X_T,s_T)_j
=
\frac{\operatorname{Cov}_{w,T}(X_{T,j},s_T)}
{\operatorname{Var}_{w,T}(s_T)},
\label{eq:haufe_general}
\end{equation}
provided that $\operatorname{Var}_{w,T}(s_T)>0$. Weighted covariance and variance use weights $w_i/\sum_{i\in T}w_i$. Coordinate $j$ is therefore the weighted least-squares slope obtained by regressing feature $j$ on the fitted score. The pattern describes covariance with a model prediction; it is neither a causal effect nor a derivative of strength with respect to the underlying geometry.

For interpretation of the frame-invariant directional model, let $X_T^{\mathrm{dir}}$ be the 729-coordinate $\mathcal B_0^z$ block normalized by ROI size and define
\begin{equation}
a_T^{\mathrm{dir}}=\mathcal H_T(X_T^{\mathrm{dir}},s_T^{\mathrm{dir}}).
\label{eq:haufe_directional}
\end{equation}
For the later signed distance $H_0$ matched pilot, the same transformation is applied instead to the signed distance $H_0$ persistence image block; that distinct pattern is denoted $a_T^{\mathrm{SD}}$. Thus $a_T^{\mathrm{dir}}$ and $a_T^{\mathrm{SD}}$ use the same mathematical transformation but are patterns in different predictor spaces and should not be conflated.

For the invariant directional analysis, patterns with nonzero norm were normalized as $\widetilde a_T^{\mathrm{dir}}=a_T^{\mathrm{dir}}/\|a_T^{\mathrm{dir}}\|_2$, and the \emph{consensus Haufe pattern} was the mean of these normalized patterns from the outer fits on coordinates available across fits. All invariant fits retained the same 728 nonterminal coordinates. Pairwise cosine similarity, sign agreement, and connected regions on the parameter lattice summarized stability. The sign rule required activity and a common sign in at least 12 of 13 fits. Coordinates in the upper quartile of consensus amplitude were used only for visualization; complete diagnostics are given in Appendix~\ref{app:invariant_haufe}. Because the outer training sets overlap, these stability summaries are dependent.
\end{revision}

\subsection{Equal removal perturbations}

Each baseline was modified by removing
\[
r_i=\lfloor0.02|B_i|+0.5\rfloor
\]
bone voxels. Diffuse removal was eligible throughout the ROI except for the first and last $0.50~\mathrm{mm}$. Localized removal was further restricted to a central $1.50~\mathrm{mm}$ band. Within each eligible set, voxels were ranked by distance to marrow, followed by a deterministic hash and voxel index. Selecting the first $r_i$ voxels gave removal sets $U_i^D$ and $U_i^L$. The diffuse structure is
\[
B_i^D=B_i\setminus U_i^D.
\]
The localized structure is
\[
B_i^L=B_i\setminus U_i^L.
\]
Since $|U_i^D|=|U_i^L|=r_i$, both structures have identical BV/TV in the same ROI. This fixes retained bone amount while allowing its spatial arrangement and other morphometric coordinates to differ. Acceptance required a six-connected bone component spanning the loading direction. Appendix~\ref{app:mechanics} gives the deterministic construction.

For coarsening factor $q$, a coarse element $Q_e$ retained fractional bone occupancy
\[
\rho_e(B)=\frac{1}{|Q_e|}\sum_{x\in Q_e}\mathbf1_B(x).
\]
Its element stiffness was
\[
K_e(B)=\rho_e(B)K_{s,e},
\]
where $K_{s,e}$ is the stiffness matrix of a solid element with the same dimensions. Equilibrium minimizes
\[
\mathcal U_B(u)=\frac12\sum_eu_e^{\mathsf T}K_e(B)u_e
\]
subject to prescribed axial compression. The primary calculation used $q=2$, trilinear hexahedra with full Gauss integration, solid Young's modulus of $2000~\mathrm{MPa}$, Poisson ratio of $0.25$, and axial strain $\varepsilon=0.001$. Lateral displacements were free except for constraints removing rigid motion. No occupancy threshold was applied. Factor $q=3$ was a specified resolution sensitivity \cite{vanRietbergen1995,vanRietbergen1996,Chevalier2007}.

The apparent modulus was $\mathcal E_q(B)= F/(A_i\varepsilon)$, where $F$ is the mean magnitude of the top and bottom axial reactions and $A_i$ is the cross-sectional area of the original digital ROI. This elastic quantity differs from measured failure strength $y_i$. Scaling the common tissue modulus scales reactions and apparent moduli equally, and therefore cancels from relative contrasts within each specimen.

Writing $E_{X,i}^{(q)}=\mathcal E_q(B_i^X)$ for $X\in\{B,D,L\}$, with $B_i^B=B_i$, define the relative change for $X\in\{D,L\}$ by
\[
R_{X,i}^{(q)}=\frac{E_{X,i}^{(q)}-E_{B,i}^{(q)}}{E_{B,i}^{(q)}}.
\]
The localized minus diffuse contrast is
\[
C_i^{(q)}=\frac{E_{L,i}^{(q)}-E_{D,i}^{(q)}}{E_{B,i}^{(q)}}.
\]
The contrast with equal weight per validation group was $\widehat C^{(q)}=G^{-1}\sum_iw_iC_i^{(q)}$. A negative value favors lower stiffness after localized removal. Its interval used 200,000 group bootstrap resamples. Solver residuals, reaction balance, and energy consistency were checked independently; complete tolerances and resolution handling are given in Appendix~\ref{app:mechanics}.

\subsection{Testing scalar topological interpretations}

For each removal, the relative persistence image displacement was
\[
d_{k,i}^{X,B}=\frac{\|p_{k,i}^X-p_{k,i}^B\|_2}{\|p_{k,i}^B\|_2}.
\]
The corresponding mechanical loss was
\[
L_{X,i}^{(q)}=1-\frac{E_{X,i}^{(q)}}{E_{B,i}^{(q)}}.
\]
These quantities test the specific interpretation that greater global image displacement indicates greater elastic damage. Persistence image grids and weights were unchanged from baseline.

\begin{revision}
A second post hoc pilot selected the core nearest the cohort median BV/TV without using strength or topology. From 120 deterministic candidates with equal bone removal, admissible pairs had identical BV/TV and differences below $0.10$ cohort standard deviations in each remaining morphometric coordinate. A regional score then selected the admissible pair with the greatest signed $H_0$ separation. Its source was the \emph{signed distance $H_0$ Haufe pattern} $a^{\mathrm{SD}}$, obtained by applying Equation~\eqref{eq:haufe_general} to the human $H_0$ persistence image model rather than to the directional predictors. Each candidate image was projected onto normalized $a^{\mathrm{SD}}$ weights in a fixed $5\times5$ region. The region and selection rule were specified before the mechanical calculations. The hypothesis was that the higher regional score would correspond to higher simulated stiffness. Because the score came from the same human strength analysis, this tests an interpretation informed by those outcomes rather than external prediction. Appendix~\ref{app:pilot} records the regional bounds, matching criteria, and selection rule.
\end{revision}

\section{Results}
\label{sec:results}

\subsection{Signed distance prediction and the original benchmark}

Dimension zero had the lowest observed error among the signed distance representations (Table~\ref{tab:prediction_main}). Its RMSE with equal weight per validation group was $1.639~\mathrm{MPa}$, compared with $2.012~\mathrm{MPa}$ for $H_1$ and $2.240~\mathrm{MPa}$ for $H_2$. The corresponding point differences from morphometry and BV/TV were $-0.362$ and $-0.596~\mathrm{MPa}$. Removing each group's fixed prediction errors in turn left the $H_0$ minus morphometry difference negative in all 13 deletions, ranging from $-0.450$ to $-0.219~\mathrm{MPa}$. This check describes sensitivity of the observed errors; no model was refitted.

\begin{table}[htbp]
\centering
\caption{Original prediction results with equal weight per validation group. Errors and their intervals are in MPa. Intervals resample fixed outer prediction errors. Asterisks identify exploratory models for individual homology dimensions.}
\label{tab:prediction_main}
\begin{tabularx}{\linewidth}{Yrrrr}
\toprule
Predictors & RMSE & 95\% interval & MAE & $Q_G^2$\\
\midrule
Training mean & 2.455 & [1.570, 3.307] & 1.938 & $-0.155$\\
BV/TV & 2.235 & [1.193, 3.112] & 1.675 & 0.042\\
Morphometry $m$ & 2.001 & [1.284, 2.700] & 1.492 & 0.232\\
$H_0$ persistence image$^{\ast}$ & 1.639 & [1.078, 2.243] & 1.254 & 0.485\\
$H_1$ persistence image$^{\ast}$ & 2.012 & [1.255, 2.799] & 1.603 & 0.224\\
$H_2$ persistence image$^{\ast}$ & 2.240 & [1.430, 3.051] & 1.767 & 0.038\\
Combined persistence image $p$ & 1.952 & [1.182, 2.697] & 1.512 & 0.270\\
$m\oplus p$ & 2.030 & [1.329, 2.723] & 1.610 & 0.210\\
\bottomrule
\end{tabularx}
\end{table}

The combined persistence image model did not establish an advantage over morphometry. Its paired difference was $\widehat\Delta_{\mathrm{rep}}=-0.049~\mathrm{MPa}$, with 95\% interval $[-0.491,0.393]~\mathrm{MPa}$. Adding the combined image to morphometry gave $\widehat\Delta_{\mathrm{add}}=+0.028~\mathrm{MPa}$, with interval $[-0.252,0.341]~\mathrm{MPa}$. Both intervals include improvements and deteriorations. The lower exploratory $H_0$ error therefore motivates a $H_0$ hypothesis without replacing the original primary result.

\subsection{Directional frame dependence and invariant Betti tensors}
\label{sec:legacy_directional_results}
\label{sec:directional_results}

The reference directional filtration gave a different ordering: RMSE for the loading direction was $1.789~\mathrm{MPa}$ for $\mathcal B_1^z$, $2.007~\mathrm{MPa}$ for $\mathcal B_0^z$, and $2.248~\mathrm{MPa}$ for $\mathcal B_2^z$. However, holding the physical direction at $z$ and exchanging the transverse coordinate in $(C_z,P_y,P_z)$ for that in $(C_z,P_x,P_z)$ changed the representation. The median correlation within specimens between first Betti tensors was 0.978, and their median relative $L^1$ difference was 0.128. The alternative frame gave an RMSE of $1.871~\mathrm{MPa}$. Smoothing the covariance neighborhood did not remove the dependence (Appendix~\ref{app:directional_audit}).

Under the invariant filtration, $H_0$ had the lowest error for the loading direction (Table~\ref{tab:directional_betti}). The paired $\mathcal B_0^z$ minus $\mathcal B_1^z$ difference was $-0.267~\mathrm{MPa}$, with interval $[-0.389,-0.149]~\mathrm{MPa}$. Relative to reference $\mathcal B_0^z$, the invariant model changed RMSE by $-0.398~\mathrm{MPa}$, with interval $[-0.673,-0.029]~\mathrm{MPa}$. This comparison concerns the complete reformulation, which changed covariance weighting and orientation coordinates as well as removing frame dependence; it does not isolate invariance as the sole cause of the error reduction.

\begin{table}[htbp]
\centering
\caption{Post hoc invariant directional prediction for the loading direction $z$. Each individual tensor contains 729 coordinates before filtering.}
\label{tab:directional_betti}
\begin{tabularx}{\linewidth}{Yrrr}
\toprule
Representation & RMSE (MPa) & MAE (MPa) & $Q_G^2$\\
\midrule
$\mathcal B_0^z$ & 1.610 & 1.214 & 0.504\\
Euler tensor $\mathcal E^z$ & 1.688 & 1.254 & 0.454\\
$\mathcal B_1^z$ & 1.877 & 1.532 & 0.326\\
$\mathcal B_0^z\oplus\mathcal B_1^z\oplus\mathcal B_2^z$ & 1.817 & 1.431 & 0.368\\
$\mathcal B_2^z$ & 2.666 & 2.207 & $-0.360$\\
\bottomrule
\end{tabularx}
\end{table}

For invariant $H_0$, the errors for the individual directions were $2.006$, $1.944$, and $1.610~\mathrm{MPa}$ for $x$, $y$, and $z$. The $z$ minus $x$ interval excluded zero, but those for $z$ versus $y$ and the transverse mean did not. The transverse mean gave $1.893~\mathrm{MPa}$, concatenating all axes gave $1.791~\mathrm{MPa}$, and the axial minus transverse contrast gave $2.231~\mathrm{MPa}$. These results favor the loading axis by point estimate without establishing a general advantage over both transverse directions.

\subsection{Comparison of $H_0$ representations and structural descriptors}
\label{sec:h0_b0_compare}

For a direct comparison, the prespecified $30\times20$ signed distance $H_0$ image was rerun through the later directional prediction pipeline. Its RMSE was $1.630~\mathrm{MPa}$, a change of $0.009~\mathrm{MPa}$ from the original $1.639~\mathrm{MPa}$. The paired invariant $\mathcal B_0^z$ minus signed distance $H_0$ difference was $-0.021~\mathrm{MPa}$, with interval $[-0.234,0.210]~\mathrm{MPa}$. The available errors do not distinguish their predictive performance.

The two representations induced related cohort geometries: linear CKA was 0.639 and Spearman correlation between pairwise distances was 0.622. Concatenation gave an RMSE of $1.580~\mathrm{MPa}$, but its paired intervals against signed distance $H_0$ and invariant $\mathcal B_0^z$ both included zero (Appendix~\ref{app:directional_audit}). Thus representational differences were observed without a resolved incremental prediction benefit.

Table~\ref{tab:comparator_summary} places the later structural descriptors alongside their contemporaneous morphometry reference, whose RMSE was $2.007~\mathrm{MPa}$. None of these additions established improvement over that reference. The organization proxy gave the lowest augmented point estimate, $1.918~\mathrm{MPa}$, but its paired interval included zero. These analyses used the combined persistence image in their additional comparisons, rather than a new direct evaluation against $H_0$. They therefore do not establish superiority of $H_0$ over every structural descriptor. Complete results, including small increments over the combined image and sensitivities in the FX subset, remain in Appendix~\ref{app:comparators}.

\begin{table}[htbp]
\centering
\caption{Later structural comparisons. The final two columns give the paired change after adding each block to contemporaneous morphometry. All entries are in MPa. These results are post hoc and are not paired contrasts against the original $H_0$ model.}
\label{tab:comparator_summary}
\begin{tabularx}{\linewidth}{Yrrr}
\toprule
Added descriptor block & Standalone RMSE & $\Delta$RMSE & 95\% interval\\
\midrule
Skeleton & 2.163 & $+0.105$ & $[-0.241,0.568]$\\
Axial graph & 2.507 & $+0.604$ & $[0.183,1.035]$\\
Organization proxy & 2.458 & $-0.0889$ & $[-0.3793,0.1335]$\\
Local orientation & 2.222 & $+0.349$ & $[-0.040,0.745]$\\
Surface tensor $W_1^{0,2}$ & 2.695 & $+0.299$ & $[-0.069,0.734]$\\
\bottomrule
\end{tabularx}
\end{table}

\begin{figure}[htbp]
\centering
\includegraphics[width=0.94\linewidth]{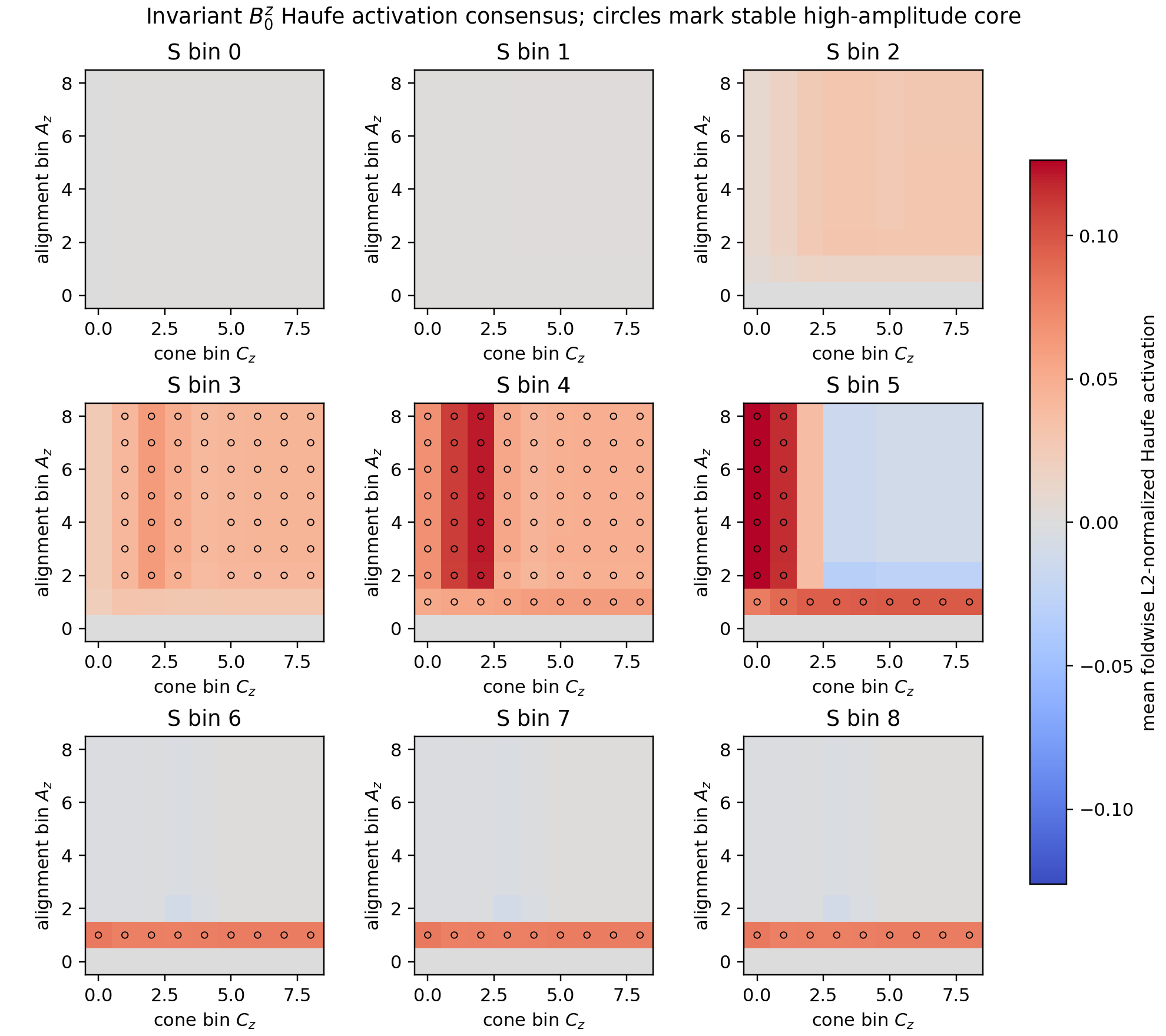}
\caption{\textcolor{revisioncolor}{Consensus directional Haufe activation pattern for invariant $\mathcal B_0^z$. Panels fix the principal axis degeneracy coordinate $S$ and display cone deficit $C_z$ against directional covariance deficit $A_z$. Values average the patterns from the outer fits after normalization to unit norm $\widetilde a_T^{\mathrm{dir}}$. Circles mark the descriptive stable region of high amplitude. All bin indices are relative to each specimen.}}
\label{fig:b0_haufe_consensus}
\end{figure}

\begin{revision}
\subsection{Stability of the invariant $H_0$ Haufe pattern}
\label{sec:b0_haufe_results}

All 13 outer invariant $\mathcal B_0^z$ fits retained the same 728 coordinates; the terminal coordinate was excluded because its raw count was constant. The directional Haufe activation patterns normalized to unit norm $\widetilde a_T^{\mathrm{dir}}$ had median pairwise cosine similarity 0.996 and minimum 0.980. The corresponding ridge coefficient maps on the original feature scale had median and minimum similarities of 0.977 and 0.819. Removing one Haufe pattern from the outer fits at a time left the cosine similarity of the consensus pattern above 0.99995. These are stability diagnostics across overlapping training sets, not independent replications.

Of 729 coordinates, 471 were active and positive in at least 12 fits, and 219 were active and negative in at least 12. Positive coordinates contributed 94.1\% of the stable absolute consensus mass. The descriptive high amplitude region comprised 176 positive coordinates in one connected region of the parameter lattice (Figure~\ref{fig:b0_haufe_consensus}). This mass was concentrated at particular relative levels: 43.8\% in cone bins 0--2, 25.4\% in covariance deficit bin 1, and 72.6\% in degeneracy bins 3--5.
\end{revision}

\begin{revision}
Higher fitted strength scores therefore covaried predominantly with larger normalized component counts over a broad set of relative filtration levels. This statement concerns the sign and magnitude of the directional Haufe pattern at sampled tensor coordinates; it does not imply that the same connected components persist across the region or that those coordinates identify mechanical load paths. Adding voxel spacing changed RMSE from $1.610$ to $1.612~\mathrm{MPa}$, with paired difference $+0.002~\mathrm{MPa}$ and interval $[-0.014,0.023]~\mathrm{MPa}$. This diagnostic found little incremental value from an additive spacing term but does not rule out resolution effects in the representation.
\end{revision}

\subsection{Elastic consequences of equal bone removal}

Every core produced accepted diffuse and localized variants with identical retained bone counts. The largest six-connected component contained at least 99.54\% of remaining bone. Localized variants retained a median $5.786~\mu\mathrm{m}$ greater mean thickness than diffuse variants, confirming that equal BV/TV did not match every geometric property.

All 72 primary mechanical calculations passed the residual, reaction, and energy criteria, with one run requiring the allowed tighter solver tolerance. Both removals reduced apparent modulus in every core. At $q=2$, localized removal produced lower modulus in 19 of 24 cores and in 11 of 13 group means. Its additional mean loss was 0.536 percentage points of baseline modulus, with interval $[0.281,0.796]$ percentage points (Table~\ref{tab:mechanics_primary}).

\begin{table}[htbp]
\centering
\caption{Mechanical changes as percentages of baseline modulus, with equal weight assigned to each validation group. A negative contrast $C=R_L-R_D$ indicates lower stiffness after localized removal. Intervals use 200,000 group bootstrap resamples.}
\label{tab:mechanics_primary}
\begin{tabularx}{\linewidth}{Yrr}
\toprule
Quantity & Factor two & Factor 3\\
\midrule
Diffuse change $R_D$ & $-2.978$ & $-2.806$\\
95\% interval & $[-3.152,-2.779]$ & $[-2.952,-2.639]$\\
Localized change $R_L$ & $-3.515$ & $-3.312$\\
95\% interval & $[-3.830,-3.197]$ & $[-3.590,-3.032]$\\
Localized minus diffuse $C$ & $-0.536$ & $-0.507$\\
95\% interval & $[-0.796,-0.281]$ & $[-0.751,-0.266]$\\
\bottomrule
\end{tabularx}
\end{table}

Factor 3 preserved the sign of the localized minus diffuse contrast in every core. The contrast across validation groups was $-0.507\%$, and the correlation between contrasts at the two scales was 0.9986. Absolute modulus remained resolution dependent: the ratio of the group mean baseline modulus at factor 3 to that at factor 2 was 1.086. The sensitivity analysis supports consistency of the paired ordering at these two discretizations, rather than convergence of absolute modulus.

\subsection{Limits of scalar topological interpretations}

Diffuse removal produced the larger persistence image displacement in dimensions zero and one for every core, that is, for $k \in \{0, 1\}$ and $i = 1,\dots,24$,
\begin{equation}
d_{k,i}^{D,B}>d_{k,i}^{L,B}.
\label{eq:distanceordering}
\end{equation}
Median $H_0$ distances were 0.342 for diffuse removal and 0.106 for localized removal. In the 19 cores with greater localized mechanical loss, the ordering of $L_{X,i}^{(2)}$ was therefore opposite to that of $d_{0,i}^{X,B}$. A nondecreasing scalar function mapping this global persistence image distance to elastic loss cannot reproduce both variants of any such pair. This is a statement about the specified distance and model, rather than about all information in the persistence image.

The separate pilot also failed to support the proposed score ordering. The selected pair had identical BV/TV of 0.234178, with every other morphometric difference below $0.10$ cohort standard deviations and joint standardized norm 0.0945. Its $H_0$ regional scores were 8.4688 and 9.5436. The member with the higher score had slightly lower modulus at both resolutions: $102.220$ versus $102.615~\mathrm{MPa}$ at factor 2, and $114.546$ versus $114.894~\mathrm{MPa}$ at factor 3. Relative high minus low contrasts were $-0.385\%$ and $-0.303\%$. This selected pair contradicts the proposed monotone transfer of that regional score to the simulated modulus; it does not evaluate the original strength predictor on an independent specimen.

\section{Discussion}

\subsection{Geometric contribution and empirical scope}

The principal methodological contribution is the removal of an arbitrary transverse coordinate from a directional topological representation. Once the physical direction is fixed, changing the basis used to describe its perpendicular plane should not change a descriptor conditioned on that axis. The reference construction did change under this operation. The proposed coordinates depend instead on cone occupancy, a covariance quadratic form, and an eigenvalue ratio. Their invariance follows directly from Euclidean geometry and orthogonal similarity, and propagates to the sampled cubical Betti tensors. This property holds independently of the strength results.

The empirical contribution is more limited. Dimension zero had the lowest observed error within both the signed distance and invariant directional constructions, with similar errors in the harmonized comparison. This convergence identifies a focused hypothesis about organization of connected components. It is not an independent replication: the invariant construction was developed after the check of the reference construction and earlier outcomes were known. Furthermore, the original combined persistence image comparison did not establish improvement over morphometry. Retaining these distinctions allows the mathematical result and the exploratory biomedical evidence to be assessed on their respective grounds.

The redesign also changed Gaussian weighting and the information encoded by the orientation coordinates. Consequently, the improvement over the reference directional model cannot be attributed solely to removing frame dependence. A prospective comparison with controlled ablations would be needed to separate these contributions. Even without such a performance claim, avoiding an irrelevant coordinate choice is a useful design requirement for computational representations intended to be compared across image acquisitions.

\subsection{Interpretation of $H_0$ in the two filtrations}

\begin{revision}
Section~\ref{sec:sedt_persistence} gives the formal interpretation of signed distance $H_0$. For the present results, the important point is that an $H_0$ interval records both the depth at which an interior component appears and the scale at which it merges as the sublevel complex grows. Mean trabecular thickness summarizes diameters of maximal inscribed spheres and therefore need not encode the same sequence of connectivity changes. The two descriptors can reflect related local scale without being mathematically equivalent.
\end{revision}

The interpretation depends on the entire interval. An $H_0$ class dying at a positive threshold merges only after marrow has entered the filtered complex. The regional score used in the pilot included births of approximately $-754$ to $-581~\mu\mathrm{m}$ and persistences of $506$ to $675~\mu\mathrm{m}$, permitting death values on either side of zero. Calling that region a direct thickness measurement would therefore omit part of its filtration meaning. Likewise, $H_2$ describes cavities of the evolving complex, which need not be anatomical cavities of the original segmentation. Its weaker prediction in this cohort does not imply general mechanical irrelevance.

\begin{revision}
Invariant directional $\mathcal B_0^d$ asks a different question. At each parameter triple it counts components in a complex selected by directional occupancy, covariance concentration, and principal axis degeneracy. The lower-dimensional cells can inherit their coordinatewise minima from different incident voxels. Thus these components belong to the defined cubical multifiltration and need not be components of a simultaneously thresholded voxel mask. The tensor records the number present at each sampled level; unlike the signed distance persistence module, it does not retain maps identifying which classes survive between levels. A stable positive region of the directional Haufe pattern consequently supports an association with component counts across several constraints, rather than a claim about particular components that persist over long parameter ranges or mechanical load paths.
\end{revision}

The moderate agreement between the two cohort geometries is consistent with their distinct constructions. However, different representations can encode overlapping predictive information. The combined $H_0$ model had the lowest point error in its comparison, while both paired intervals included zero. Its usefulness as an augmented predictor therefore remains undetermined.

\subsection{Mechanical experiments and interpretation}

\begin{revision}
Prediction across specimens and response to an intervention are different mathematical relationships. The first estimates $y$ from a representation of $B$ within the observed cohort. The second compares a mechanical functional at $B$ and at a deliberately altered structure. No implication between these relationships follows from a low prediction error alone. In particular, the regional score derived from the signed distance Haufe pattern used in the pilot projects new geometry onto a Haufe pattern estimated from the human strength model; treating that projection as a mechanical ordering introduces an additional transfer hypothesis \cite{Haufe2014}.
\end{revision}

\begin{revision}
The equal removal experiment supplies a controlled test of one such interpretation. Bone amount was fixed exactly between variants, yet localized removal caused greater stiffness loss in most cores despite smaller global $H_0$ and $H_1$ image displacements. The additional loss was small in magnitude but consistent at the two discretizations. The matched pilot narrowed conventional morphometric differences further and again placed the higher topological score below the lower score in elastic modulus. Together, these observations limit two proposed scalar surrogates: global persistence image distance and the regional score derived from the signed distance $H_0$ Haufe pattern.
\end{revision}

These findings concern simulated linear elastic stiffness. The supervised endpoint was measured failure strength, which can also depend on yielding, buckling, damage, and material heterogeneity. Those mechanisms were not represented in the mechanical calculation. The perturbations therefore constrain the interpretation of the strength associations without refuting their observed predictive value. They also do not test the invariant directional representation under intervention, because the reported counterfactual scores came from signed distance persistence.

\subsection{Other structural descriptors and biomedical scope}

The structural comparators represent complementary geometric ideas: a thinned network, axial graph routes, distributions of local plate and rod orientations, spatial misorientation, and surface normal anisotropy. Their inclusion broadens the geometric comparison, but their tested implementations have specific limits. Graph conductance omitted the areas of trabecular cross sections and constitutive behavior; the organization proxy did not reconstruct individual trabeculae; and the surface tensor block retained eigenvalue shape while excluding magnitude and principal direction. Their errors cannot be generalized to every method in those families.

The modest or unresolved increments in these post hoc analyses show why geometric novelty and predictive benefit require separate evidence. They also leave a specific comparison unfinished: the focal $H_0$ models have not been evaluated prospectively against all structural blocks in one common pipeline. The present results are therefore most relevant to computational characterization of excised trabecular architecture. They do not establish a tool for clinical fracture risk or performance at routine clinical CT resolution.

\subsection{Limitations and validation}

\begin{revision}
The cohort contains 24 cores and only 13 assigned validation groups, with strong imbalance among clinical categories. Because those groups are not verified donors, residual donor overlap across folds cannot be excluded. The small sample also limits adjustment for confounding and the precision of calibration. Bootstrap intervals condition on fixed outer predictions, while the training sets used for Haufe pattern stability overlap. Neither procedure accounts fully for repeating data collection, representation development, and model selection.
\end{revision}

\begin{revision}
Outcome blinding protected the initial image construction, but the persistence image grids used pooled baseline diagrams from all specimens. This outcome-independent use of held-out image distributions differs from a wholly representation estimated only from training data. The later emphasis on $H_0$, invariant coordinate construction, and Haufe pattern summaries adds selection uncertainty. Their exploratory intervals should therefore guide future hypotheses rather than serve as confirmatory tests after repeated comparisons.
\end{revision}

Image conventions introduce further limits. The ROI samples part of the mechanically tested core; distance is measured between voxel centers; and segmentation determines the interface used by several descriptors. The rectangular connectivity correction does not explicitly correct the curved ROI wall, and anisotropy was measured within an inscribed prism. For directional tensors, binning performed separately for each specimen means that a shared index denotes a relative level, not a common numerical threshold. Frame invariance also does not remove errors from physically rotating and resampling a voxel lattice.

Finally, deterministic voxel removal is not biological remodeling, and coarse elasticity with fractional occupancy does not establish mechanics at native resolution. The two coarsening factors preserved the reported orderings while changing absolute modulus. The pilot matched in morphometry examined one deliberately selected pair, with small stiffness differences. These experiments support conclusions only within their stated geometries and constitutive assumptions.

The next validation should use independent specimens with verified donor identities and a prespecified analysis specification. Signed distance $H_0$, invariant directional $\mathcal B_0^z$, their combination, and the intended morphometric comparators should be fixed before outcome access. That specification must include ROI and segmentation rules, physical grids and weights, directional normalization and binning, preprocessing, tuning, and the primary comparison. Such a study would test whether the observed association based on component counts transfers beyond this development cohort.

\section{Conclusions}

The proposed directional filtration removes ambiguity in the transverse frame through geometrically defined coordinates and an explicit cubical extension. In this cohort, dimension zero had the lowest observed error within both signed distance and invariant directional topology, but the later comparison did not distinguish their predictive performance or establish a benefit from combining them. The original combined persistence image benchmark also did not establish improvement over morphometry. Equal removal experiments and a pair matched in morphometry limited the interpretation of signed distance scores as monotone surrogates for elastic stiffness. These results support external evaluation of the prespecified $H_0$ representations while preserving the distinction between a mathematically defined geometric descriptor, an observed strength association, and a mechanical response to intervention.

\section*{Data and code availability}

The software implementations used in this study are publicly available on GitHub. The persistence-image computation is available in \href{https://github.com/jhnrckmnznrs/persistence_images}{\texttt{persistence\_images}}, the trabecular bone morphometry Rust implementation in \href{https://github.com/jhnrckmnznrs/bone_morphometry}{\texttt{bone\_morphometry}}, and the directional Betti and frame-invariant directional filtration Python implementation in \href{https://github.com/jhnrckmnznrs/directional_betti}{\texttt{directional\_betti}}. The source human cohort is described by Jin et al.\ and Jin \cite{Jin2017,Jin_2016}.

\section*{Funding}

This project has received funding from the European Union Horizon Europe research and innovation program under the Marie Sk\l{}odowska-Curie Actions grant agreement No. 101120290 (GAP).

\section*{Acknowledgements}

John Rick Manzanares and Pawe\l{} D\l{}otko acknowledge support from the Dioscuri Program and the hosting of the Institute of Mathematics of the Polish Academy of Sciences.

\clearpage
\appendix
\setcounter{table}{0}
\setcounter{figure}{0}
\renewcommand{\thetable}{A\arabic{table}}
\renewcommand{\thefigure}{A\arabic{figure}}
\renewcommand{\theHtable}{appendix.\arabic{table}}
\renewcommand{\theHfigure}{appendix.\arabic{figure}}
\section{Image construction and numerical representation}
\label{app:images}

This appendix records implementation details that are not needed for the main mathematical development. Numerical choices are retained from the completed study, and later prediction runs are distinguished from the original benchmark.

\subsection{Image coordinates and physical spacing}
The volumes were unsigned 16-bit TIFF stacks with array axes ordered as slice, row, and column, denoted by $Z,Y,X$. Specimen spacing was read from TIFF and ImageJ metadata \cite{imagej2021}. It was effectively isotropic within each scan. Transverse image widths were 6.740--6.978~mm and reconstructed $Z$ extents were 10.400--11.906~mm. These reconstructed extents were not assumed to be the mechanical gauge lengths.

\subsection{Region of interest}

The interval along the $Z$-axis containing the specimen was first identified from the reconstructed image stack. For each axial slice $z$, we computed the population standard deviation of its voxel intensities,
\[
s_i(z)=\operatorname{sd}_{y,x} I_i(z,y,x),
\]
where $i$ denotes the specimen. This gives the finite profile $(s_i(z))_{z=0}^{N_{z,i}-1}$ along $Z$. Slices containing mostly exterior background generally have little intensity variation, whereas slices intersecting the specimen contain a mixture of bone, marrow, and structural boundaries and therefore tend to have larger values of $s_i(z)$.

Small fluctuations between adjacent slices were reduced using a centered moving median over seven slices. The window size was determined from a target physical width of $0.15$ mm by dividing this width by the axial voxel spacing of each scan, rounding to the nearest number of slices, and increasing an even result to the next odd integer. This procedure gave a seven-slice window for every specimen.

For each specimen, a single Otsu threshold was estimated from the lower half of the smoothed axial profile, defined as the profile values no greater than that specimen's median. This threshold was then applied to the entire smoothed profile. Slices with values above the threshold were marked as candidate specimen-bearing slices.

Short runs of slices classified as background but lying between specimen-bearing slices were filled when their physical length did not exceed \(0.10\) mm. This prevented isolated fluctuations in the axial profile from splitting an otherwise continuous specimen interval.

To exclude irregularities associated with the physical cut surfaces, $0.50$ mm was removed from each detected end. An analysis interval with a nominal height of $8.50$ mm was then centered within the remaining specimen. Since the boundaries of this interval had to coincide with voxel boundaries, its realized physical height ranged from $8.4822$ to $8.5000$ mm across specimens.

In the transverse plane, the ROI was a circle of radius 2.75~mm centered on the geometric center of the image. A voxel belonged to the resulting cylinder when the center of that voxel lay within the circle. Radii of 2.50, 2.75, and 3.00~mm were compared during the process. A radius of 2.75~mm was the largest of these candidates that remained within every usable core, and the same physical radius was therefore applied to all scans. Consequently, voxels outside the cylinder could be excluded from calculations rather than treated as marrow. The resulting cylindrical voxel set is denoted by $R_i\subset\Omega_i$.

\subsection{Segmentation details}

The smoothing and thresholding equations are given in Section~\ref{sec:image_construction}. Here, only implementation details not stated there are recorded. The Gaussian kernel had standard deviation $0.50$ voxel and radius two voxels. Otsu's threshold was computed from a 256-bin histogram of the smoothed intensities inside the ROI using scikit-image \cite{Otsu1979,vanderWalt2014}. Voxels outside the ROI were stored as zero in the binary image, but the ROI mask continued to exclude them from the marrow phase.

Local Otsu alternatives were inspected during image construction, but the single threshold for each specimen was retained because it was spatially consistent and introduced no additional neighborhood parameter. No component filtering, hole filling, morphological operation, manual correction, or further ROI revision followed segmentation. Such operations would alter the topology being measured. The resulting BV/TV ranged from 0.121 to 0.431.

\subsection{Morphometric conventions}

BV/TV, Tb.Th, Tb.Sp, and Conn.D were calculated within the cylindrical ROI using BoneJ \cite{Hildebrand1997,BoneJ2021}. The conventional edge correction for a rectangular image stack was retained for connectivity. It corrects contributions at the image faces but does not explicitly account for the curved cylindrical wall, which is reported as a limitation. The five variables in $m_i$ were fixed before strength modelling; additional BoneJ summaries were not screened against strength. Thickness and separation standard deviations were omitted to limit correlated predictors, and Euler characteristic was not added separately because it already enters connectivity density.

Degree of anisotropy was estimated by the mean intercept length method \cite{ODGAARD1997315}. The calculation used 2048 directions, 64 lines per direction, a sampling increment of $\sqrt{3}$ voxels, and five repetitions. To prevent voxels outside the cylinder from being interpreted as marrow, this calculation used the square prism inscribed in the cylindrical ROI over its full height. The mean degree of anisotropy across repetitions was used. The acute angle between the principal direction and the loading axis was retained only for a specified sensitivity analysis.

\subsection{Persistence image discretization}
Cubical persistence was computed using GUDHI \cite{gudhi:urm,Kaczynski2004,Bleile2022}. A separate Rust implementation evaluated the Gaussian cell integrals in Equation~\eqref{eq:pi}. This was the integrated persistence image representation used for prediction; a grid of sampled Gaussian surface values was not substituted.

The dimension-specific grid sizes and weight saturation constants are reported in Table~\ref{tab:pi_grids} in the main text.
The saturation constants and grid choices were fixed using pooled baseline diagrams before outcome access. The recorded analysis reported agreement with the next finer grid; the numerical tolerance and complete physical grid edges are part of the fixed representation specification. Baseline domains, cell edges, and weights were reused for all counterfactuals. Bandwidths of 50, 75, and 100~$\mu$m were sensitivity choices around the principal 25-$\mu$m value. Changing bandwidth changes the represented surface, whereas changing the grid refines its integration.

\FloatBarrier
\section{Additional prediction results}
\label{app:prediction}

\subsection{Metrics and paired comparisons}
For the outer predictions and group weights defined in Equation~\eqref{eq:risk}, mean absolute error was
\[
\mathrm{MAE}_G=\frac1G\sum_iw_i|y_i-\widehat y_i|.
\]
The coefficient of determination was
\[
Q_G^2=1-\frac{\sum_iw_i(y_i-\widehat y_i)^2}
{\sum_iw_i(y_i-\bar y_w)^2}.
\]
Here the weighted observed mean is
\[
\bar y_w=\frac1G\sum_iw_i y_i.
\]
The denominator of $Q_G^2$ uses the pooled observed mean. It is not the error of the training-mean predictor fitted separately within each fold. A negative value means the pooled predictions have greater squared error than that pooled mean. Metrics and correlations weighted by individual cores were descriptive.

Table~\ref{tab:paired_comparisons} retains all original primary paired comparisons. Reported paired contrasts are retained at their original precision; subtracting rounded model errors can differ in the final decimal. The fraction of bootstrap samples favoring a model is descriptive and is not a posterior probability or a conventional significance level.

\begin{table}[htbp]
\centering
\caption{Original paired RMSE comparisons. $M$ denotes morphometry and $P$ combined persistence images. Negative differences favor the candidate. $b_-$ is the fraction of bootstrap samples with a negative difference. Units are MPa.}
\label{tab:paired_comparisons}
\begin{tabular}{llrrr}
\toprule
Reference & Candidate & $\Delta$RMSE & 95\% interval & $b_-$\\
\midrule
BV/TV & $M$ & $-0.234$ & $[-0.704,0.327]$ & 0.795\\
$M$ & $P$ & $-0.049$ & $[-0.491,0.393]$ & 0.613\\
$M$ & $M\oplus P$ & $+0.028$ & $[-0.252,0.341]$ & 0.434\\
\bottomrule
\end{tabular}
\end{table}

The leave-one-group deletion check reported with Table~\ref{tab:prediction_main} used the already obtained outer errors and did not refit any model; it is therefore a sensitivity analysis of the observed error difference rather than an additional validation experiment.

\subsection{Specified sensitivity analyses}

Adding the angle between the principal trabecular direction and the loading axis to morphometry produced an RMSE of 1.967~MPa and $Q^2=0.258$. Adding this angle to the model containing both morphometry and topology produced an RMSE of 1.938~MPa and $Q^2=0.280$. These small changes in the estimates did not alter the primary conclusion.

The 25~$\mu$m Gaussian bandwidth remained the principal persistence image choice specified before outcomes were examined. The models combining all three homology dimensions had RMSE values of 2.246, 2.409, and 2.491~MPa at bandwidths of 50, 75, and 100~$\mu$m, respectively. Their corresponding $Q^2$ values were 0.033, $-0.112$, and $-0.190$. These results were reported as sensitivity analyses and were not used to choose another bandwidth.

\subsection{Calibration and residual checks}
Calibration was described by $y_i=a+b\widehat y_i+\varepsilon_i$, using equal aggregate weight per group. The $H_0$ slope was 0.929 with interval $[0.378,1.440]$. Mean observed minus predicted strength was 0.167~MPa with interval $[-0.584,1.003]$~MPa. The observation--prediction Pearson correlation was 0.702, and the correlation between observed strength and residual was 0.657 with interval $[0.112,0.871]$. The latter describes an increasing residual trend across observed strength; because the residual contains the observation, it is not an independent diagnostic of mechanism. No recalibration was fitted and evaluated on these same observations.

Residual checks included age, sex, clinical group, voxel spacing, image dimensions, and grayscale summaries. The largest absolute correlation among the principal models and auxiliary variables was 0.339 between age and the residual of morphometry plus topology, with interval $[-0.184,0.699]$. All other reported intervals also included zero. The small sample and imprecise intervals cannot rule out confounding. Clinical and sex-specific summaries were descriptive, and no residual check was used to select predictors.

\FloatBarrier
\section{Directional multifiltration}
\label{app:directional_audit}

\subsection{Reference Betti tensors and transverse frame check}

The reference hard neighborhood construction in Equation~\eqref{eq:legacy_directional_filters} documents the directional analysis preceding the invariant reformulation. For the prespecified $z$ direction its errors with equal weight per validation group were 1.748~MPa for the Euler tensor, 2.007~MPa for $\mathcal B_0^z$, 1.789~MPa for $\mathcal B_1^z$, 2.248~MPa for $\mathcal B_2^z$, 1.749~MPa for $-\mathcal B_1^z+\mathcal B_2^z$, and 1.940~MPa when all three Betti tensors were entered as separate blocks. The paired $\mathcal B_1^z-\mathcal B_0^z$ RMSE difference was $-0.218$~MPa with interval $[-0.358,-0.041]$~MPa, whereas $\mathcal B_2^z-\mathcal B_1^z$ was $+0.458$~MPa with interval $[0.081,0.926]$~MPa.

For the transverse frame check with the physical direction fixed at $z$, the standard coordinate choice $(C_z,P_y,P_z)$ and the choice after a $90^\circ$ roll $(C_z,P_x,P_z)$ had median $\mathcal B_1$ correlation within specimens 0.978 and median relative $L^1$ difference 0.128. Their nested model RMSEs were 1.789 and 1.871~MPa, respectively; the paired difference between the rolled and standard choices was $+0.082$~MPa with interval $[-0.058,0.287]$~MPa. This check identifies a definitional dependence even though the present cohort does not resolve which roll predicts better.

The Gaussian covariance of Equation~\eqref{eq:gaussian_weight} was then substituted without changing the legacy coordinate selection. The median roll correlation became 0.936 and the median relative $L^1$ difference 0.153. Outcome-blind global measures of transverse orientation imbalance were strongly associated with this roll sensitivity, whereas deviation of the effective weighted support radius from its cohort median was not. The Gaussian nested model also exhibited one severe extrapolation in one outer fold caused by inner selection of a very small ridge penalty. A diagnostic with a fixed penalty removed most of the apparent roll difference, but that diagnostic was post hoc and is not used as a replacement result. Together these checks motivate changing the coordinate definition itself rather than tuning the legacy roll.

\subsection{Exactness and frame invariance}

The frame-invariant implementation was checked independently at the filtration and topology levels. For synthetic data, a $90^\circ$ lattice preserving roll about fixed $z$ produced identical invariant Betti tensors, as predicted by Proposition~\ref{prop:frameinvariance}. Exact Betti computation was compared with explicit homology computed from the explicit chain complex on random small cubical complexes. For every one of the 24 cohort specimens and each direction $x,y,z$, all $9^3$ tensor entries satisfied nonnegativity and
\[
\mathcal E^d=\mathcal B_0^d-\mathcal B_1^d+\mathcal B_2^d
\]
exactly, with terminal state $(1,0,0,1)$. The terminal state and nonnegativity are implementation checks. Since $\mathcal B_1$ is recovered from the Euler relation, satisfaction of that relation is not by itself an independent homology validation; the explicit chain-complex comparison provides the independent check.

\subsection{Paired comparisons for invariant $H_0$}

\begin{table}[htbp]
\centering
\caption{Selected paired group bootstrap comparisons in the frame-invariant analysis. Differences are candidate minus reference RMSE; negative values favor the candidate. The signed distance $H_0$ row uses the harmonized rerun described in Section~\ref{sec:h0_b0_compare}.}
\label{tab:invariant_paired}
\begin{tabular}{llrr}
\toprule
Reference & Candidate & $\Delta$RMSE (MPa) & 95\% interval\\
\midrule
Training mean & $\mathcal B_0^z$ & $-0.847$ & $[-1.338,-0.292]$\\
BV/TV & $\mathcal B_0^z$ & $-0.631$ & $[-1.434,0.103]$\\
Signed distance $H_0$ & $\mathcal B_0^z$ & $-0.021$ & $[-0.234,0.210]$\\
$\mathcal B_1^z$ & $\mathcal B_0^z$ & $-0.267$ & $[-0.389,-0.149]$\\
$\mathcal B_0^x$ & $\mathcal B_0^z$ & $-0.396$ & $[-0.647,-0.039]$\\
$\mathcal B_0^y$ & $\mathcal B_0^z$ & $-0.334$ & $[-0.647,0.109]$\\
Transverse mean $\mathcal B_0$ & $\mathcal B_0^z$ & $-0.283$ & $[-0.542,0.095]$\\
Signed distance $H_0$ & $H_0\oplus\mathcal B_0^z$ & $-0.050$ & $[-0.248,0.131]$\\
$\mathcal B_0^z$ & $H_0\oplus\mathcal B_0^z$ & $-0.030$ & $[-0.101,0.020]$\\
\bottomrule
\end{tabular}
\end{table}

The invariant $\mathcal B_0^z$ model also improved on the legacy dependent on the coordinate frame $\mathcal B_0^z$ point estimate from 2.007 to 1.610~MPa. Using the same group resamples, the difference between the invariant and reference constructions was $-0.398$~MPa with interval $[-0.673,-0.029]$~MPa. Because the invariant construction was developed after the legacy result, this is a methodological comparison within this cohort rather than an external validation.

\begin{revision}
\subsection{Invariant $H_0$ Haufe pattern stability}
\label{app:invariant_haufe}

The stability analysis in Section~\ref{sec:b0_haufe_methods} refitted every outer $\mathcal B_0^z$ model using the already selected penalty and reproduced the previously recorded out-of-fold predictions to maximum absolute difference $4.4\times10^{-16}$. Every fit retained 728 coordinates. The terminal coordinate was the only coordinate removed by the within each fold variance filter applied to the raw counts. Penalty selection had already chosen only $\alpha=562.341325$ or $1000$ across the 13 outer fits.

Pairwise cosine similarity of the normalized to unit norm directional Haufe patterns $\widetilde a_T^{\mathrm{dir}}$ had median 0.9961, minimum 0.9803, and maximum 0.9998. Raw ridge coefficient maps on the original feature scale were less stable, with median 0.9773 and minimum 0.8194. The consensus itself was insensitive to any single outer fit: consensus maps obtained after omitting one fit had cosine at least 0.999958 with the complete consensus. Of the 729 coordinates, 471 were positive in at least 12 outer fits and 219 were negative in at least 12, conditional on the coordinate being active in at least 12 fits. The stable positive regions contained 94.1\% of stable absolute consensus mass.

The descriptive upper-quartile stable core contained 176 coordinates and formed one positive region connected by six-neighbor adjacency. Its ranges of relative bin indices were $C_z=0,\ldots,8$, $A_z=1,\ldots,8$, and $S=3,\ldots,8$, with center weighted by consensus pattern magnitude $(3.35,3.83,4.55)$. The range itself is broad, so the marginal distribution of consensus-pattern mass is more informative: 43.8\% of total absolute consensus mass occurred in cone bins 0--2, 25.4\% in alignment bin 1, and 72.6\% in shape bins 3--5. A much smaller stable negative region was concentrated principally at moderate and high relative shape levels. These indices cannot be translated into one common physical threshold because each specimen and direction is rescaled by its own coordinate extrema before binning.

\begin{figure}[htbp]
\centering
\includegraphics[width=0.92\linewidth]{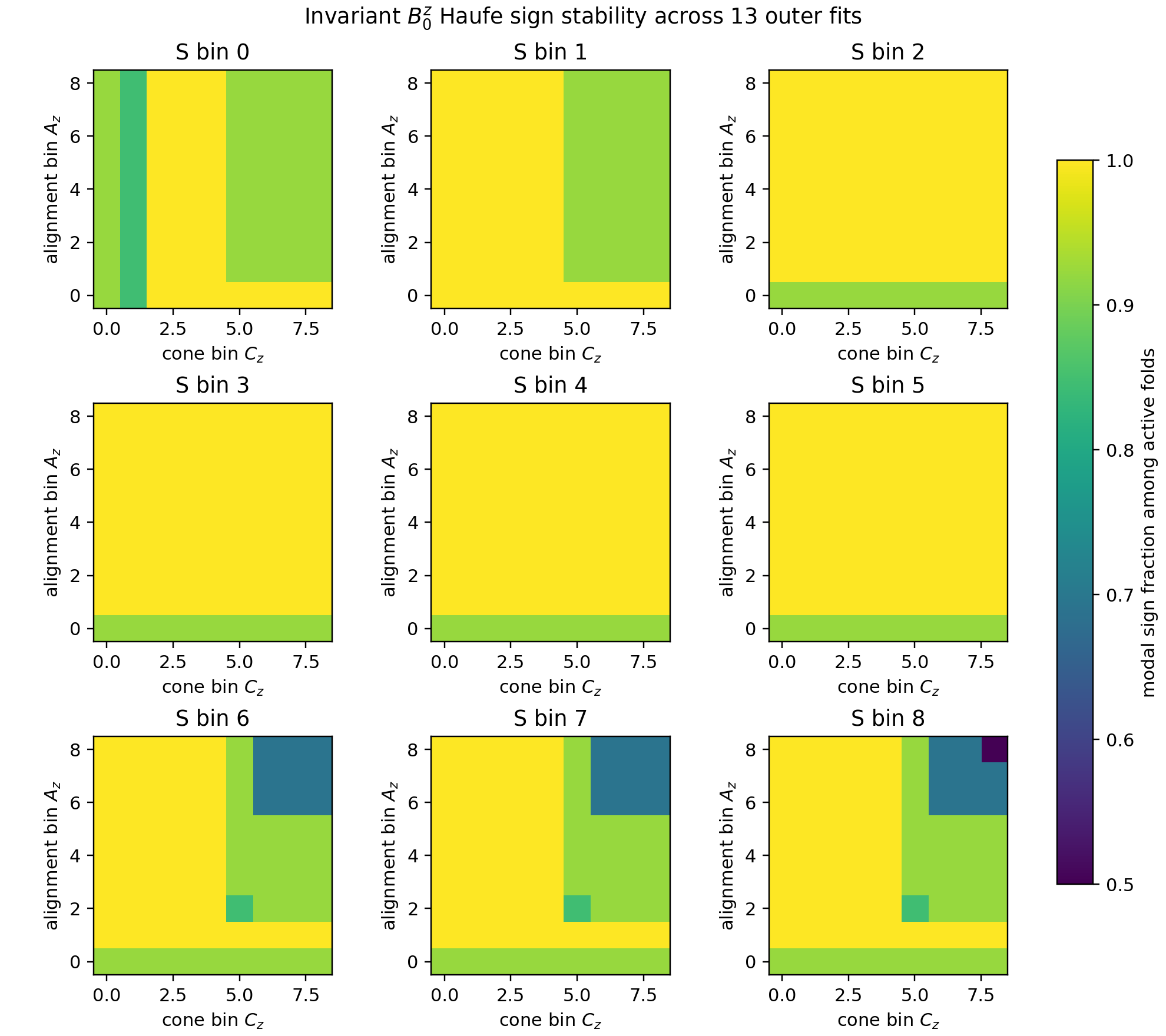}
\caption{Sign stability of the frame-invariant $\mathcal B_0^z$ directional Haufe activation pattern across the 13 outer fits. Each panel fixes one relative shape bin. Values are the fraction of active outer fits agreeing with the modal sign at that coordinate. Outer training sets overlap, so this is a stability diagnostic rather than independent replication.}
\label{fig:b0_haufe_stability}
\end{figure}

A post hoc nuisance diagnostic added voxel spacing as a separate scalar block to the $\mathcal B_0^z$ model. RMSE changed from 1.6097 to 1.6120~MPa; the paired candidate minus reference difference was $+0.0023$~MPa with 95\% group bootstrap interval $[-0.0144,0.0227]$~MPa. Adding both BV/TV and spacing gave RMSE 1.6123~MPa. These tests do not prove resolution invariance, but they do not support a simple additive spacing explanation of the $\mathcal B_0^z$ result.
\end{revision}

\begin{figure}[htbp]
\centering
\includegraphics[width=0.92\linewidth]{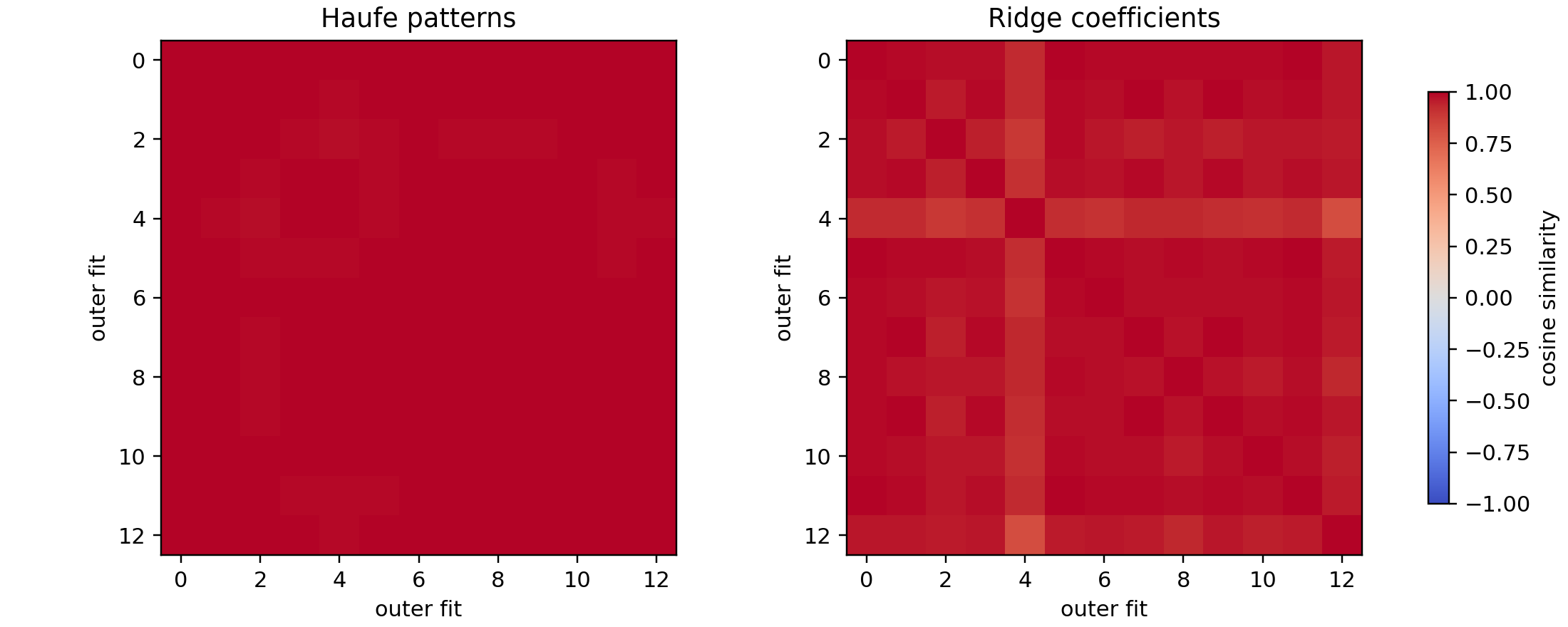}
\caption{Cosine similarity between outer fit maps for frame-invariant $\mathcal B_0^z$. Haufe activation patterns are more stable across outer fits than the corresponding ridge coefficient maps on the original feature scale.}
\label{fig:b0_haufe_coef_compare}
\end{figure}

\section{Post hoc structural comparisons}
\label{app:comparators}

These structural comparisons were developed after the original strength results were known. Each block was specified before it was linked to the strength outcomes, but all comparisons remain post hoc. The later reference errors were 2.007~MPa for morphometry and 1.989~MPa for the combined persistence image, compared with 2.001 and 1.952~MPa in the original analysis. The 0.037-MPa persistence image difference met the recorded 0.05-MPa tolerance. Every incremental contrast below uses the contemporaneous reference from the same run.

\subsection{Skeleton graph coordinates}

Each binary bone mask was converted to a three-dimensional skeleton one voxel in width using Lee's topology-preserving thinning. Skeleton voxels were classified as endpoints, slabs, or junction voxels from their degree in the 26-neighborhood. Connected junction voxels were grouped into junction vertices, and branches were traversed using the same stateful rules as BoneJ, beginning at endpoints and then proceeding through junctions. No cycle pruning was applied. Branch lengths were evaluated in physical units using the isotropic voxel spacing for each specimen; mean branch tortuosity was the branch length divided by the straight endpoint distance for branches with a positive endpoint distance.

Let $V_i$ be the physical ROI volume in mm$^3$, $b_i$ be the number of graph branches, $j_i$ be the number of junction clusters, $e_i$ be the number of endpoint voxels, $q_i$ be the graph cycle rank, $c_i$ be the number of graph components, and $\bar\ell_i$ be the mean branch length in $\mu$m. The eight skeleton predictors were
\[
\left(\frac{b_i\bar\ell_i/1000}{V_i},\frac{b_i}{V_i},\frac{j_i}{V_i},\frac{e_i}{V_i},\frac{q_i}{V_i},\frac{c_i}{V_i},\bar\ell_i,\bar\tau_i\right),
\]
where $\bar\tau_i$ is the mean branch tortuosity. The first coordinate is total graph branch length per ROI volume in mm/mm$^3$. The next five are branch, junction cluster, endpoint, cycle, and graph-component densities per mm$^3$. The spanning indicator was constant across all 24 specimens.

For a compressed graph with vertex set $V$, branch set $E$, and $c$ components, the graph cycle rank is $|E|-|V|+c$. This graph quantity is distinguished from the homology of the full voxel filtration.

\subsection{Axial graph coordinates}

The eight skeleton variables were centered and standardized using group weights within each training fold, and their block was divided by the square root of the active dimension. Nested ridge regression used the 49-value penalty grid from the main analysis, with numerical ties resolved toward the larger penalty. The primary skeleton comparison added this block to morphometry. The 19 FX cores formed a specified secondary sensitivity analysis.

A subsequent comparison added the fixed combined $H_0\oplus H_1\oplus H_2$ persistence image to the skeleton block. The reciprocal addition and the comparison conditional on morphometry were secondary. No image dimension, bandwidth, region, or skeleton coordinate was reselected. Scalar predictions had to reproduce the preceding run, and the combined persistence image error had to remain within 0.05~MPa of its original value.

As a further extension of the validated centerline graph, five mechanically oriented descriptors were constructed without access to apparent strength, clinical group, or validation group labels. Let $G=(V,E)$ be the compressed multigraph and let $L_e$ be the physical length of branch $e$. Axial terminal sets $T_0$ and $T_1$ contained graph vertices with at least one constituent skeleton voxel within one voxel of the minimum or maximum occupied skeleton $Z$ coordinate, respectively. The one-voxel cap was chosen during geometric quality control because an extreme occupied slice can be reached through a slab voxel of degree two without creating an explicit compressed graph vertex. If $H_z$ is the physical separation between the minimum and maximum occupied skeleton slices, axial shortest-path tortuosity was
\[
\tau_z=\frac{1}{H_z}\min_{P:T_0\leadsto T_1}\sum_{e\in P}L_e.
\]
Axial edge connectivity $\lambda_z$ was the maximum number of paths from $T_0$ to $T_1$ that shared no graph branches when every AnalyzeSkeleton branch had unit capacity and the artificial terminal connectors had capacity greater than the total number of branches.

To describe how much of the graph could participate in axial connection, the graph was augmented by auxiliary source and sink vertices and decomposed into biconnected blocks. Let $E_z\subseteq E$ contain the branches assigned to blocks on the source-to-sink path of the resulting block-cut tree. The path participation length fraction was
\[
C_z=\frac{\sum_{e\in E_z}L_e}{\sum_{e\in E}L_e}.
\]
Self-loops and dangling blocks outside this path in the block-cut tree were not counted as participating. A separate graph transport surrogate assigned branch conductance $g_e=1/L_e$, fixed terminal potentials to one and zero, and solved Kirchhoff balance at all interior vertices. If $G_z$ is the resulting total source current, normalized axial conductance was
\[
\widetilde G_z=H_zG_z.
\]
For branch current $I_e$, dissipation was $d_e=I_e^2L_e$. The fifth descriptor was the Gini coefficient of the branch dissipation values,
\[
D_z=\frac{2\sum_{r=1}^{m}r\,d_{(r)}}{m\sum_{r=1}^{m}d_{(r)}}-\frac{m+1}{m},
\]
where $d_{(1)}\le\cdots\le d_{(m)}$. The conductance and dissipation quantities are graph transport surrogates; they are not finite element estimates and do not incorporate trabecular cross-sectional area or tissue constitutive properties.

The block $(\tau_z,\lambda_z,C_z,\widetilde G_z,D_z)$ was specified before it was linked to the strength outcomes. Bridge burden, articulation burden, the length carrying nonzero current, and summaries of concentrated dissipation remained descriptive after a redundancy check that did not use outcomes. The primary comparison added the block to the skeleton; secondary comparisons added it to morphometry, combined persistence images, or skeleton plus persistence images. The sensitivity analysis restricted to the FX cores retained the same definitions.

Finally, representation overlap was examined without reading strength, clinical group, or validation group labels. The eight skeleton variables were compared with $H_0$, $H_1$, $H_2$, and their concatenation using linear CKA and Spearman correlation between the upper triangles of the Euclidean distance matrices between specimens. For each persistence image we also calculated total mass, Euclidean norm, birth centroid, persistence centroid, and the fraction of mass at negative birth values, using the prespecified physical birth and persistence domains. Correlations of these summaries with skeleton descriptors were descriptive. \begin{revision}
A separate comparison projected each human $H_0$ persistence image onto the already fixed $5\times5$ region of the signed distance $H_0$ Haufe pattern used in the mechanistic pilot. Because this pattern was derived from the original human strength model, this comparison interprets a pattern derived from those outcomes rather than providing independent validation.
\end{revision}

\subsection{Plate and rod organization proxy}

A later structural comparator was an open proxy for the structural organization index of Cox et al. \cite{Cox2026SOI}. The published SOI is defined from orientations of individual trabecular plates and rods obtained with ITS \cite{Liu2011ITS}; the present estimator does not reconstruct individual trabeculae and is therefore not described as ITS or as literal Cox SOI. Instead, the already validated topology-preserving skeleton was used only as sampling support. At each sampled skeleton voxel, a local diameter proxy was defined as twice the Euclidean distance transform value. This proxy based on center distance is not the maximal-sphere local thickness used by BoneJ. Foreground voxels within a sphere of radius 2.5 times the proxy were used to form a covariance matrix. If the square roots of its ordered eigenvalues were $a\le b\le c$, local shape was summarized by
\[
E_{\mathrm{local}}=\frac{a}{b}-\frac{b}{c}.
\]
Samples with $E_{\mathrm{local}}<-0.20$ were classified as plate-like and represented by the shortest covariance axis, interpreted as the local plate normal. Samples with $E_{\mathrm{local}}>0.20$ were classified as rod-like and represented by the longest covariance axis. Intermediate samples were not used in the orientation densities. The neighbourhood multiplier and classification threshold were fixed using engineered geometries only. Across 152 synthetic rod and plate cases spanning oblique orientations, thicknesses, subvoxel offsets, and surface perturbations, the largest recovered error in the unoriented axis was $0.171^\circ$, and all tested support points received the intended rod or plate class. The $4.5^\circ$ orientation criterion used for this gate followed the engineered lattice validation reported during development of SOI \cite{Cox2022SOI}.

For the organization calculation, plate normal and rod axis directions were mapped to polar angle $\theta\in(-90^\circ,90^\circ)$ and elevation $\phi\in(0^\circ,90^\circ)$. Separate reflected Gaussian kernel density estimates $f_P$ and $f_R$ were evaluated on a $0.5^\circ$ grid with bandwidth $4.5^\circ$. Working in radians, the uniform density over this angular rectangle is
\[
C_u=\frac{1}{\pi(\pi/2)}.
\]
Let $V_{\max}=0.954077$ denote the reference integral for a maximally concentrated interior orientation under the same kernel and grid. Plate organization was
\[
pO=\frac{\iint \max(f_P-C_u,0)\,d\theta\,d\phi}{V_{\max}}.
\]
Rod organization was
\[
rO=\frac{\iint \max(f_R-C_u,0)\,d\theta\,d\phi}{V_{\max}}.
\]
The plate--rod orientation overlap was
\[
prO=1-\frac{1}{2}\iint |f_P-f_R|\,d\theta\,d\phi.
\]
The resulting scalar was
\[
\mathrm{SOI}_{\mathrm{proxy}}=pO\,rO\,prO.
\]
The numerical implementation reproduced $C_u=0.202642$ and the reference $V_{\max}$ before human application.

Because local skeleton sampling can give long structures more observations than short structures, the human calculation included a separate sampling sensitivity analysis that did not use outcomes. For each specimen, let $t_{50}$ be the median of this local diameter proxy over the previously defined skeleton. One representative skeleton voxel was retained per physical cube of side $t_{50}$, chosen as the voxel centre nearest the cube center, with deterministic tie breaking; this $1.0\,t_{50}$ rule was fixed as the primary sampling scheme. Sensitivity schemes used cube sides $0.5\,t_{50}$, $1.5\,t_{50}$, and $2.0\,t_{50}$. All 24 specimens retained at least 25 plate and 25 rod samples under every scheme, the maximum range of the SOI proxy within a specimen across schemes was 0.0406, and the weakest cohort Spearman correlation between a sensitivity scheme and the primary scheme was 0.890. The primary $1.0\,t_{50}$ SOI proxy was therefore retained as the only predictor from this branch; $pO$, $rO$, $prO$, and analogues using elevation alone were retained only for description.

The organization proxy used the same nested prediction and bootstrap procedure as the preceding comparators. Its primary addition was to morphometry; secondary additions were to DA, combined persistence images, or their combination with morphometry. The refit restricted to the FX cores retained the same definition. No orientation component, sampling density, neighborhood multiplier, or classification threshold was reselected after the strength outcomes were linked.

In these formulas $prO$ measures overlap of angular densities on the specified rectangle. This numerical overlap is distinct from the physical angle between a plate normal and a rod axis at neighboring locations. Also, a uniform density in $(\theta,\phi)$ coordinates is not a uniform distribution of axes with respect to spherical area. These conventions are part of the defined proxy; they should not be replaced silently by a different orientation statistic.

\subsection{Local orientation variation}

To test whether local spatial variation of trabecular orientation explained information retained by the other representations, a later analysis reused the plate normal and rod axis samples. This analysis was specified without using strength outcomes. For two unoriented axes $u$ and $v$, same type misorientation was
\[
\theta(u,v)=\arccos\!\left(|u^{\mathsf T}v|\right),
\]
which is invariant to reversing either axis. For a plate normal $n$ and rod axis $a$, departure from physical plate--rod orthogonality was
\[
\delta_{\perp}(n,a)=\arcsin\!\left(|n^{\mathsf T}a|\right),
\]
so that $\delta_{\perp}=0$ when the rod lies in the plate plane. Neighbours were defined within a physical radius $2t_{50}$, where $t_{50}$ is the median local diameter proxy from the organization sampling analysis. Same-type disagreement was averaged first over neighbours at each retained location and then across locations. Plate--rod disagreement was calculated as the equal mean of plate centered and rod centered local means, preventing a class with more pair combinations from dominating by count alone.

The primary sampling grid was $1.0\,t_{50}$. The three predictors were mean local plate normal misorientation, rod axis misorientation, and plate--rod orthogonality deviation. Sensitivity grids at $0.5\,t_{50}$ and $1.5\,t_{50}$ required at least 25 locally supported centers and Spearman correlation at least 0.85 with the primary descriptors. The $2.0\,t_{50}$ grid was a sparsity stress test because one specimen lacked sufficient rod support. The selected block was added primarily to morphometry and secondarily to the organization proxy, combined persistence images, and morphometry plus persistence images. Models restricted to the FX cores used the same definitions.

\subsection{Minkowski surface tensors}

As a separate surface geometric comparator, we evaluated Minkowski tensors of rank two that are invariant under translation, following the integral geometric framework used for trabecular bone by Callens et al. \cite{Callens2021Minkowski}. Under the convention used here,
\[
W_1^{0,2}=\frac{1}{3}\int_{\partial K} n\otimes n\,dA.
\]
The tensor weighted by curvature is
\[
W_2^{0,2}=\frac{1}{3}\int_{\partial K} H\,n\otimes n\,dA.
\]
Here $n$ is the surface unit normal and $H$ is mean curvature. The discrete tensor kernel was first validated on analytic cuboids for translation invariance, rotation covariance, scale laws, and alternate triangulations. Human surfaces were then extracted from the previously calibrated Gaussian smoothed signed distance field at 20, 30, and 40~$\mu$m, with 30~$\mu$m primary. To avoid manufacturing a cylindrical bone/ROI wall, the ROI selected support rather than cutting the surface. A common 160-$\mu$m interior margin, equal to four times the largest sensitivity bandwidth, was used at all three scales; $W_1^{0,2}$ support was selected by triangle barycentres and $W_2^{0,2}$ support by shared edge midpoints.

For each tensor, eigenvalues were converted to two dimensionless shape quantities. For the positive semidefinite $W_1^{0,2}$ tensor with $0\le\lambda_1\le\lambda_2\le\lambda_3$, define
\[
A_{102}=1-\frac{\lambda_1}{\lambda_3}.
\]
The second shape quantity is
\[
E_{102}=\frac{\lambda_2}{\lambda_3}.
\]
The same normalized eigenvalue shape construction, using absolute eigenvalues, was evaluated for $W_2^{0,2}$ before the strength outcomes were examined. At that stage, the $W_2^{0,2}$ quantities weighted by curvature failed the prespecified criterion for stability under smoothing: their maximum 20/30/40-$\mu$m ranges within a specimen reached 0.603 and 0.466, exceeding the limit of 0.15. In contrast, the smallest Spearman correlation across the sensitivity scales was 0.997 for $A_{102}$ and $E_{102}$, and their maximum ranges were 0.0288 and 0.0185. The prespecified stability criterion therefore retained only $(A_{102},E_{102})$ at 30~$\mu$m as the $W_1^{0,2}$ block of two variables; $W_2^{0,2}$ was not taken forward to strength modelling.

The primary tensor comparison added the retained $W_1^{0,2}$ block to morphometry. Secondary comparisons added it to DA, the organization proxy, combined persistence images, and morphometry plus persistence images, using the same nested ridge procedure. Tensor magnitudes, angles of the principal axes, alternative bandwidths, and $W_2^{0,2}$ were not selected after the strength outcomes were examined.

\subsection{Skeleton comparison results}

All 24 specimens passed skeleton extraction and structural checks. Skeleton ridge regression gave RMSE 2.163~MPa and $Q_G^2=0.105$. Morphometry gave 2.007~MPa, and morphometry plus skeleton gave 2.113~MPa. The paired addition was $+0.105$~MPa (95\% interval $[-0.241,0.568]$~MPa), with negative differences in 0.298 of bootstrap resamples. For the FX cores, the addition was $-0.022$~MPa, with interval $[-0.108,0.050]$~MPa.

The subsequent joint comparison met both criteria for reproducing the reference results. Skeleton alone gave 2.163~MPa, the combined persistence image 1.989~MPa, and their concatenation 1.981~MPa (Table~\ref{tab:skeleton_ph_posthoc}). The combined image estimate differed by 0.037~MPa from its original result.

\begin{table}[htbp]
\centering
\caption{Later comparison using the validated skeleton graph and the previously specified persistent homology representation. RMSE gives equal weight to each validation group. These models do not replace the original primary morphometry versus topology comparison.}
\label{tab:skeleton_ph_posthoc}
\begin{tabular}{lrr}
\toprule
Model & RMSE (MPa) & $Q^2$ \\
\midrule
Morphometry ridge & 2.007 & 0.229 \\
Skeleton ridge & 2.163 & 0.105 \\
Persistent homology ridge & 1.989 & 0.243 \\
Morphometry + skeleton & 2.113 & 0.146 \\
Morphometry + persistent homology & 1.988 & 0.244 \\
Skeleton + persistent homology & 1.981 & 0.249 \\
Morphometry + skeleton + persistent homology & 1.982 & 0.249 \\
\bottomrule
\end{tabular}
\end{table}

For the comparison fixed as primary within this post hoc stage, adding persistent homology to skeleton changed RMSE by $-0.182$~MPa, with a 95\% group bootstrap interval from $-0.592$ to $0.130$~MPa and bootstrap fraction 0.870 that the combined model had lower RMSE. For the full cohort, the point estimate favored persistent homology beyond skeleton, but its interval included zero. The reciprocal increment was much smaller: adding skeleton to persistent homology changed RMSE by $-0.008$~MPa, with interval $[-0.027,0.00047]$~MPa. In the specified sensitivity analysis restricted to the FX cores, adding persistent homology to skeleton changed RMSE by $-0.127$~MPa with interval $[-0.232,-0.013]$~MPa, whereas adding skeleton to persistent homology changed RMSE by only $+0.00019$~MPa. The FX result is a secondary sensitivity and was not used to redefine the comparison for the full cohort.

Outcome-independent representation analysis showed only moderate overlap between the two structural descriptions. Linear centered-kernel alignment between the skeleton block and the combined persistence image representation was 0.678, and the Spearman correlation between their specimen pair distance matrices was 0.656. Dimension-specific kernel alignment was 0.565 for $H_0$, 0.698 for $H_1$, and 0.496 for $H_2$. Some global persistence image quantities were strongly related to graph density: $H_0$ total persistence image mass correlated with cycle density at $\rho=0.934$, and the corresponding $H_1$ value was $\rho=0.914$. In contrast, several summaries describing where topology occurred in physical filtration coordinates were less closely represented by the skeleton. The strongest skeleton association with the $H_2$ birth centroid had $|\rho|=0.375$, and that with the $H_0$ persistence centroid had $|\rho|=0.556$.

\begin{revision}
The regional score derived from the fixed signed distance $H_0$ pattern was also only moderately associated with individual skeleton descriptors: its largest Spearman correlation was 0.660 with endpoint density. By contrast, unweighted mass in the same $H_0$ region correlated 0.934 with junction cluster density, 0.924 with skeleton length density, and 0.921 with branch density. These descriptive results suggest that simple topological amount is substantially shared with graph density, whereas the distribution of signal across physical birth and persistence coordinates is less completely represented by the centerline graph.
\end{revision}

The graph implementation was checked independently of these strength comparisons. On six human pilot skeletons, its Rust implementation reproduced all 25,632 AnalyzeSkeleton branch pairings and physical lengths exactly. This validates the intended graph conventions on those tested inputs.

\subsection{Axial graph comparison results}

All 24 specimens passed the graph verification checks before the five mechanically oriented descriptors were linked to outcomes. Across the cohort, axial edge connectivity ranged from 1 to 22, $C_z$ from 0.695 to 0.931, shortest-path tortuosity from 1.423 to 1.770, normalized conductance from 3.67 to 31.32, and dissipation Gini from 0.706 to 0.957. The mechanical graph block alone gave an RMSE with equal weight per validation group of 2.507~MPa and $Q^2=-0.203$.

Adding the block of five variables to the validated skeleton did not improve prediction. Skeleton plus mechanical graph had RMSE 2.235~MPa versus 2.163~MPa for skeleton alone, giving $\Delta\mathrm{RMSE}=+0.072$~MPa with 95\% group bootstrap interval $[-0.356,0.410]$~MPa. Table~\ref{tab:mechanical_graph_posthoc} gives the prespecified incremental comparisons. For the FX cores, the corresponding increment over skeleton was $+0.035$~MPa with interval $[-0.124,0.214]$~MPa.

\begin{table}[htbp]
\centering
\caption{Increment from adding the retained axial graph block to each reference. Negative differences favor addition. Units are MPa; intervals use the group bootstrap.}
\label{tab:mechanical_graph_posthoc}
\begin{tabularx}{\linewidth}{Yrr}
\toprule
Reference & $\Delta$RMSE & 95\% interval\\
\midrule
Skeleton & $+0.072$ & $[-0.356,0.410]$\\
Morphometry & $+0.604$ & $[0.183,1.035]$\\
Persistent homology & $-0.0266$ & $[-0.0531,0.00008]$\\
Skeleton + persistent homology & $-0.0181$ & $[-0.0437,0.00117]$\\
\bottomrule
\end{tabularx}
\end{table}

The comparison conditional on morphometry was unfavorable: morphometry plus mechanical graph had RMSE 2.612~MPa versus 2.007~MPa for morphometry, with $\Delta\mathrm{RMSE}=+0.604$~MPa and an interval entirely above zero. Adding the mechanical graph block to persistent homology produced only a small change for the full cohort, from RMSE 1.989 to 1.963~MPa ($\Delta\mathrm{RMSE}=-0.0266$~MPa, 95\% interval $[-0.0531,0.00008]$~MPa). In the sensitivity analysis restricted to the FX cores, this increment was $-0.00034$~MPa with interval $[-0.00070,0.00005]$~MPa. Adding the mechanical graph block after skeleton plus persistent homology likewise changed RMSE by only $-0.0181$~MPa, with interval $[-0.0437,0.00117]$~MPa. Thus the mechanically oriented centerline summaries did not provide a stable incremental signal beyond the previously defined skeleton or persistent homology representations in this cohort.

\subsection{Organization proxy results}

All 24 specimens passed the organization sampling checks. The primary proxy ranged from 0.0545 to 0.1630, with median 0.0784. The new run reproduced the preceding reference predictions to machine precision. The proxy alone gave RMSE 2.458~MPa and $Q_G^2=-0.156$.

Adding the retained SOI proxy to morphometry reduced RMSE from 2.007 to 1.918~MPa. The primary paired difference was therefore $\Delta\mathrm{RMSE}=-0.0889$~MPa, with 95\% group bootstrap interval $[-0.379,0.133]$~MPa and bootstrap fraction 0.766 that the combined model had lower RMSE. The point estimate favored the SOI proxy, but the interval included zero. The prespecified increments are summarized in Table~\ref{tab:soi_proxy_posthoc}.

\begin{table}[htbp]
\centering
\caption{Increment from adding the retained SOI proxy to each reference. Negative differences favor addition. Units are MPa; intervals use the group bootstrap.}
\label{tab:soi_proxy_posthoc}
\begin{tabularx}{\linewidth}{Yrr}
\toprule
Reference & $\Delta$RMSE & 95\% interval\\
\midrule
Morphometry & $-0.0889$ & $[-0.3793,0.1335]$\\
DA & $-0.0365$ & $[-0.1133,0.0000002]$\\
Persistent homology & $-0.0292$ & $[-0.0557,-0.0025]$\\
Morphometry + persistent homology & $-0.0270$ & $[-0.0539,-0.0004]$\\
\bottomrule
\end{tabularx}
\end{table}

For the full cohort, the increment over persistent homology was small but had a bootstrap interval below zero: RMSE decreased from 1.989 to 1.960~MPa. A similarly small reduction was observed after morphometry plus persistent homology, from 1.988 to 1.961~MPa. These changes did not persist when the models were refitted using only the 19 FX cores. In that sensitivity analysis, adding SOI proxy to morphometry increased RMSE from 1.039 to 1.111~MPa ($\Delta=+0.0721$~MPa; interval $[-0.0144,0.1757]$~MPa), and adding it to persistent homology changed RMSE from 0.9264 to 0.9271~MPa ($\Delta=+0.00064$~MPa; interval $[0.00015,0.00146]$~MPa). The proxy showed small reductions in the point errors for the full cohort against some references, but these were not retained within FX. The post hoc comparisons do not establish general complementarity.

\subsection{Local orientation comparison results}

All 24 specimens passed the orientation stability criterion that had been specified before outcome analysis. At the primary $1.0\,t_{50}$ sampling scale, mean local plate misorientation ranged from $10.56^\circ$ to $22.75^\circ$, rod misorientation from $9.61^\circ$ to $32.88^\circ$, and local plate--rod orthogonality deviation from $13.60^\circ$ to $29.30^\circ$. Across the accepted $0.5\,t_{50}$ and $1.5\,t_{50}$ sensitivity grids, the weakest Spearman correlation with the primary representation was 0.928. The three primary variables were correlated but not identical ($\rho=0.792$--$0.854$) and were retained together as the selected block.

The orientation field block alone had RMSE 2.222~MPa and $Q^2=0.055$. Adding it to morphometry increased RMSE from 2.007 to 2.357~MPa. The primary paired difference was
\[
\Delta\mathrm{RMSE}=+0.349~\mathrm{MPa},
\]
with 95\% group bootstrap interval $[-0.040,0.745]$~MPa and bootstrap fraction 0.038 that the augmented model had lower RMSE. In the refit restricted to the FX cores, adding the field block to morphometry also increased RMSE, by 0.147~MPa with interval $[-0.143,0.481]$~MPa. For the full cohort, adding the field block to persistent homology changed RMSE only from 1.989 to 1.985~MPa ($\Delta=-0.0042$~MPa; interval $[-0.0228,0.0043]$~MPa), and the increment for the FX cores was $-0.00016$~MPa. Thus explicit local orientation disorder did not provide a stable incremental strength signal beyond morphometry or persistent homology.

\subsection{Surface tensor comparison results}

A stability analysis performed before strength modelling distinguished the two Minkowski tensor constructions. Both $W_1^{0,2}$ invariants were highly stable across the 20, 30, and 40~$\mu$m surfaces: Spearman correlations across these sensitivity scales were at least 0.9965, while the largest ranges within a specimen were 0.0288 for $A_{102}$ and 0.0185 for $E_{102}$. The corresponding normalized eigenvalue shapes from the curvature-weighted $W_2^{0,2}$ tensor failed the fixed range criterion, with maximum ranges 0.603 and 0.466 despite complete manifold and support quality control. They were therefore excluded before the tensor descriptors were linked to the strength outcomes. At the primary 30-$\mu$m scale, the fixed $A_{102}$ values ranged from 0.264 to 0.750 and $E_{102}$ from 0.477 to 0.912.

The stable $W_1^{0,2}$ block alone had RMSE 2.695~MPa and $Q^2=-0.390$. Adding it to morphometry increased RMSE from 2.007 to 2.307~MPa, giving the primary
\[
\Delta\mathrm{RMSE}=+0.299~\mathrm{MPa}
\]
with 95\% group bootstrap interval $[-0.069,0.734]$~MPa and bootstrap fraction 0.058 that the augmented model had lower RMSE. Adding $W_1^{0,2}$ to DA reduced RMSE by 0.175~MPa, but the interval $[-0.591,0.235]$~MPa included zero. For the full cohort, addition to persistent homology reduced RMSE from 1.989 to 1.961~MPa ($\Delta=-0.0284$~MPa; interval $[-0.0566,0.00049]$~MPa), whereas the increment for the FX cores was $+0.00045$~MPa. The stable tensor of surface normals therefore did not improve the primary morphometry model, and its small increment over persistent homology in the full cohort was not reproduced within FX.

\subsection{Interpretation of the structural comparators}
The descriptor families ask different geometric questions. Skeleton counts describe a network after thinning; axial graph coordinates describe routes and transport on that network; the organization proxy aggregates orientation distributions; local misorientation retains neighborhood information; and $W_1^{0,2}$ integrates surface normal directions. Their different constructions do not guarantee additional predictive information in a cohort of 13 groups.

The graph conductance model uses branch length and omits trabecular area, bending resistance, and tissue properties. The organization and local orientation estimators sample neighborhoods rather than reconstructing individual trabeculae by ITS. The surface tensor analysis retained normalized eigenvalue shape, excluding magnitude and principal-axis direction. The instability of $W_2^{0,2}$ was determined before strength analysis. These restrictions specify the tested representations; the results should not be generalized to every graph, orientation, or tensor description of bone.

Representation overlap used Equation~\eqref{eq:cka} and the upper triangle distance correlation described in the main text. These summaries quantify agreement between representations of the observed specimens; they do not identify a causal contribution to strength.

\FloatBarrier
\section{Detailed perturbation and mechanical calculations}
\label{app:mechanics}

\subsection{Removal sets and acceptance criteria}

Let $n_i=|B_i|$. The common removal count was
\[
r_i=\left\lfloor0.02n_i+0.5\right\rfloor.
\]
Let $P_i\subset B_i$ be the bone voxels in the first or last 0.50~mm of the cylinder, and let
\[
\mathcal Z_i=\left\{x\in R_i:
\left|z(x)-\frac{H_i}{2}\right|\le0.75~\mathrm{mm}\right\}
\]
be the centered 1.50-mm band, where $H_i$ is the realized ROI height. The diffuse eligible set was
\[
E_i^D=B_i\setminus P_i.
\]
The localized eligible set was
\[
E_i^L=(B_i\setminus P_i)\cap\mathcal Z_i.
\]

Surface targeting used $\delta_i(x)=d\!\left(x,R_i\setminus B_i\right)$, for $x\in B_i$, computed with physical spacing $h_i$. The exterior $\Omega_i\setminus R_i$ was not included in the marrow set. Eligible voxels were ordered lexicographically by
\[
\bigl(\delta_i(x),\,\eta_i(x),\,j_i(x)\bigr),
\]
where $\eta_i(x)$ is a deterministic hash derived from the coded specimen identifier and random seed, and $j_i(x)$ is the linear voxel index used only as a final tie breaker. The hash term removes the directional bias that results from resolving equal Euclidean distances by array order.

Let $\operatorname{first}_{r_i}(E;\prec_i)$ denote the first $r_i$ voxels of $E$ under this order. Define the diffuse removal set by
\[
U_i^D=\operatorname{first}_{r_i}(E_i^D;\prec_i).
\]
Define the localized removal set by
\[
U_i^L=\operatorname{first}_{r_i}(E_i^L;\prec_i).
\]
The corresponding diffuse structure is $B_i^D=B_i\setminus U_i^D$, and the localized structure is $B_i^L=B_i\setminus U_i^L$. Hence, $|B_i^D|=|B_i^L|=|B_i|-r_i$ and therefore
\[
\mathrm{BV/TV}(B_i^D;R_i)=\mathrm{BV/TV}(B_i^L;R_i)
\]
for every specimen. The 2\% removal fraction, 1.50-mm band, 0.50-mm end protection, and deterministic tie breaking rule were established before morphometric, topological, or mechanical outcomes of the complete perturbation set were examined.

Each generated pair was accepted only if both counterfactuals were subsets of the baseline structure, diffuse and localized thinning removed exactly $r_i$ voxels, no voxel was removed from the protected end regions, every localized removal lay inside the central band, and the remaining bone contained a six-connected component spanning the loading direction. The fraction of retained bone in the largest connected component was also recorded as a structural quality measure.

\subsection{Structural changes}

For $X\in\{B,D,L\}$, the same map
\[
\mathcal M(B_i^X)=m_i^X
\]
was applied to obtain BV/TV, Tb.Th, Tb.Sp, Conn.D, and DA. The physical spacing and DA settings were identical to those used for the original cohort. The paired localized minus diffuse contrast for feature $j$ is
\[
c_{ij}=m_{ij}^{L}-m_{ij}^{D}.
\]
The perturbations are therefore \emph{material matched} by construction, while $c_{ij}$ quantifies the remaining differences in conventional morphology.

For each baseline, diffuse, and localized structure, the same signed distance transformation, cubical filtration, and persistence image mapping were applied in dimensions 0, 1, and 2. Writing $D_{k,i}^X$ for the finite persistence diagram in dimension $k$, the sequence was
\[
B_i^X\xmapsto{\;\text{signed distance and persistence}\;}D_{k,i}^X\xmapsto{\;\mathrm{PI}_k\;}p_{k,i}^{X}.
\]
The birth and persistence domains, grid resolutions, lifetime weights, and principal Gaussian bandwidth of 25~$\mu$m were those specified for the original human analysis and were applied unchanged. The lifetime saturation constants were 374.703, 715.692, and 520.009~$\mu$m for $H_0,H_1,H_2$, respectively.

For any two structures $X,Y\in\{B,D,L\}$, the baseline scaled distance was defined by
\[
d_{k,i}^{X,Y}
=
\frac{\left\|p_{k,i}^{X}-p_{k,i}^{Y}\right\|_2}
     {\left\|p_{k,i}^{B}\right\|_2}.
\]
These distances were descriptive before mechanics quantities and were not used to alter the perturbations or the persistence image design.

\subsection{Elastic discretization and numerical checks}

For each $X\in\{B,D,L\}$, the binary structure was coarsened by a factor of two in each spatial direction. Each $2\times2\times2$ block of native image voxels was represented by one finite element, whose local bone volume fraction was taken as the fraction of its native voxels classified as bone. No majority threshold was applied. Thus, for a coarse element $e$ containing the native voxel block $Q_e$,
\[
\rho_{e,i}^X=\frac{1}{|Q_e|}\sum_{x\in Q_e}\mathbf 1_{B_i^X}(x),
\]
with $0\leq\rho_{e,i}^X\leq1$ and $|Q_e|=8$ for an ordinary factor 2 cell. A singleton axial layer contains four native voxels per cell. A value of $0$ denotes an element containing no bone, a value of $1$ denotes an element completely occupied by bone, and intermediate values preserve the fractional bone content of the corresponding native voxel block.

Empty transverse padding could be removed only when both the ROI and bone masks were empty on the complete outer plane. The loading direction was never cropped. For the three specimens with an odd number of fine $Z$ slices, all slices were retained by using paired two-slice coarse layers except for one interior singleton layer nearest the specimen center; the finite element layer height was therefore $2h_i$ for paired layers and $h_i$ for the singleton.

Each active coarse cell was a standard eight-node trilinear hexahedral element evaluated with full $2\times2\times2$ Gauss integration, rather than reduced integration, following the general voxel finite element approach used for trabecular microstructures \cite{vanRietbergen1995,vanRietbergen1996,Chevalier2007}. The reference solid phase was taken as isotropic and linearly elastic with Young's modulus $E_s=2000$~MPa and Poisson ratio $\nu=0.25$. Fractional occupancy entered linearly through
\[
K_{e,i}^X=\rho_{e,i}^X K_{s,e},
\]
where $K_{s,e}$ is the stiffness of a fully occupied cell with the same physical dimensions, including any residual axial layer. The value of $E_s$ sets the absolute modulus scale of this linear model: multiplying the common solid phase modulus by a constant multiplies $E_i^B$, $E_i^D$, and $E_i^L$ by the same constant and therefore cancels from the dimensionless relative contrasts defined below. The primary interpretation consequently concerns relative contrasts within each specimen, not the absolute apparent modulus or a measured tissue modulus.

Compression was applied along $Z$ with nominal strain $\varepsilon=0.001$. The bottom loading surface had zero axial displacement, while the top loading surface received the corresponding prescribed compression. Three lateral degrees of freedom on two bottom-surface nodes were constrained to remove rigid-body motion, while the remaining lateral degrees of freedom were left unconstrained.

Let $F_{\mathrm{top},i}^X$ and $F_{\mathrm{bottom},i}^X$ denote the axial reactions on the two loading surfaces. The compressive force used to calculate apparent stress was the mean magnitude of these reactions,
\[
F_i^X= \frac{|F_{\mathrm{top},i}^X|+|F_{\mathrm{bottom},i}^X|}{2}.
\]
If $A_i$ is the cross-sectional area of the original digital cylindrical ROI, the apparent stress was
\[
\sigma_i^X=\frac{F_i^X}{A_i}.
\]
The simulated \emph{apparent Young's modulus} was then
\[
E_i^X=\frac{\sigma_i^X}{\varepsilon}.
\]
Here, $\varepsilon$ was calculated from the imposed platen displacement divided by the full ROI gauge length $H_i$. Thus, $E_i^X$ is a simulated small-strain apparent Young's modulus of the coarse grained linear-elastic structure and is distinct from the experimentally measured apparent compressive strength used in the observational prediction analysis.

The linear system was solved by matrix-free conjugate gradients with scalar Jacobi preconditioning, a primary relative tolerance of $10^{-6}$, and at most 20,000 iterations. The Jacobi preconditioner rescales the system by its stiffness diagonal to improve convergence without changing the underlying finite element problem. Acceptance additionally required an independently recomputed relative residual at most $2\times10^{-6}$, relative top to bottom reaction imbalance at most $10^{-4}$, and relative energy inconsistency at most $2\times10^{-5}$. If an otherwise unchanged problem failed any of the conditions, the relative tolerance $5\times10^{-7}$ was allowed.

A second coarse scale sensitivity was specified before the factor 3 mechanical results were inspected. The same binary baseline, diffuse, and localized inputs were block averaged at factor 3 and solved with the same material law and boundary conditions. Transverse dimensions were made divisible by three only by removing completely empty outer padding planes. The loading direction was never cropped or padded. When the number of fine $Z$ slices was not divisible by three, triplet layers were combined with one deterministic interior residual layer of one or two fine slices nearest the specimen center, and its physical height was passed explicitly to the finite element operator. The factor 3 contrasts were analyzed with the same 13-group bootstrap and were treated only as a resolution sensitivity.

\subsection{Group comparison and sensitivity summaries}

The diffuse relative change was
\[
R_{D,i}=\frac{E_i^D-E_i^B}{E_i^B}.
\]
The localized relative change was
\[
R_{L,i}=\frac{E_i^L-E_i^B}{E_i^B}.
\]
The primary comparison of spatial support was
\[
C_i=R_{L,i}-R_{D,i}=\frac{E_i^L-E_i^D}{E_i^B}.
\]
The inequality $C_i<0$ means that localized thinning produces the larger stiffness reduction at exactly the same global bone loss. The primary point estimate gave equal weight to each of the 13 validation groups. Contrasts were first averaged over specimens within each group, and the 13 group means were then averaged. Uncertainty was estimated from 200,000 nonparametric cluster bootstrap replicates obtained by resampling the 13 groups with replacement while retaining all cores within a sampled group. The 2.5th and 97.5th percentiles formed the reported 95\% confidence interval. A cluster bootstrap weighted by the number of cores was evaluated as a sensitivity analysis. As a secondary inferential diagnostic, all $2^{13}$ sign flips of the 13 group means of $C$ were enumerated. Under this diagnostic null hypothesis, positive and negative signs of each group effect are symmetric, so every possible sign allocation is considered equally.

\subsection{Complete structural results}

All 24 baselines produced complete diffuse and localized variants satisfying the prespecified geometry criteria. For every specimen, $|B_i^D|=|B_i^L|$. The final BV/TV was therefore identical within each pair. Diffuse removal extended over 7.45--7.48~mm of the unprotected height and had slice-wise normalized entropy 0.9905--0.9995. Localized removal remained in the centered band, with realized span 1.467--1.481~mm. The common 2\% global removal represented 8.15--15.87\% of eligible bone inside the localized band. The largest six-connected component retained at least 99.54\% of the remaining bone.

For each morphometric coordinate $j$, let
\[
c_{ij}=m_{ij}^{L}-m_{ij}^{D}.
\]
The mean family values and median paired contrasts are given in Table~\ref{tab:counterfactual_morphometry}. Exact equality of BV/TV did not imply equality of all remaining coordinates. The largest systematic contrast was in Tb.Th as the localized structure was thicker than the diffuse structure by a median 5.786~$\mu$m. The corresponding paired differences in Tb.Sp, Conn.D, and DA were small relative to the natural between-specimen variation. Thus, the intervention is exactly matched in material amount, but only approximately matched in conventional morphology.

\begin{table}[htbp]
\centering
\caption{Conventional morphometry of the baseline, diffuse, and localized families. Values are means over the same 24 coded specimens.}
\label{tab:counterfactual_morphometry}
\begin{tabular}{L{3.4cm}rrrr}
\toprule
Feature & Baseline & Diffuse & Localized & Median $L-D$ \\
\midrule
BV/TV & 0.251035 & 0.246014 & 0.246014 & 0 \\
Tb.Th ($\mu$m) & 234.217 & 223.413 & 229.523 & +5.786 \\
Tb.Sp ($\mu$m) & 758.774 & 757.704 & 758.759 & +0.398 \\
Conn.D (mm$^{-3}$) & 3.9016 & 4.1778 & 4.1299 & $-0.0471$ \\
DA & 0.5554 & 0.5330 & 0.5409 & +0.00854 \\
\bottomrule
\end{tabular}
\end{table}

The topological response is measured by
\[
d_{k,i}^{X,Y}= \frac{\|p_{k,i}^{X}-p_{k,i}^{Y}\|_2}{\|p_{k,i}^{B}\|_2}.
\]
For $k=0$ and $k=1$, diffuse thinning produced a larger change from baseline than localized thinning. For all 24 specimens,
\[
d_{k,i}^{D,B}>d_{k,i}^{L,B}.
\]
The median diffuse to baseline distances were 0.342 in $H_0$ and 0.223 in $H_1$, compared with 0.106 and 0.083 for localized thinning. $H_2$ changed much less and showed no analogous ordering.

\begin{table}[htbp]
\centering
\caption{Relative persistence image distances among the counterfactual families. Each entry is median [interquartile range] over the 24 paired specimens, using the baseline norm in the denominator for all three contrasts.}
\label{tab:counterfactual_pi_distance}
\begin{tabular}{lrrr}
\toprule
Dimension & Diffuse--baseline & Localized--baseline & Localized--diffuse \\
\midrule
$H_0$ & 0.342 [0.255, 0.379] & 0.106 [0.051, 0.128] & 0.244 [0.205, 0.287] \\
$H_1$ & 0.223 [0.146, 0.269] & 0.083 [0.046, 0.102] & 0.143 [0.120, 0.174] \\
$H_2$ & 0.0101 [0.0064, 0.0182] & 0.0104 [0.0080, 0.0181] & 0.00939 [0.00630, 0.0166] \\
\bottomrule
\end{tabular}
\end{table}

The two constructions therefore satisfy exact material matching but produce distinct spatial and topological responses. These quantities were established before the mechanical contrasts were interpreted and were not used to retune the perturbations or the persistence image representation.

\subsection{Complete mechanical results}

All 72 baseline, diffuse, and localized finite element simulations passed the numerical acceptance criteria. Seventy-one were accepted at the primary conjugate-gradient tolerance of $10^{-6}$. One diffuse simulation failed only the reaction equilibrium criterion at that stopping tolerance and passed all unchanged criteria after the allowed tightening to $5\times10^{-7}$. Across the accepted simulations, the independently recomputed relative residual was at most $9.998\times10^{-7}$, the reaction equilibrium error was at most $6.834\times10^{-5}$, and the energy inconsistency was at most $8.096\times10^{-7}$.

Absolute apparent moduli varied widely across specimens, with baseline mean 146.85~MPa and median 115.52~MPa. Because the factor 2 occupancy mesh is a coarse grained discretization, the primary mechanical quantities are the paired relative contrasts rather than these absolute values. The estimates with equal weight per validation group are shown in Table~\ref{tab:counterfactual_fea}.

\begin{table}[htbp]
\centering
\caption{Paired counterfactual mechanical response. Estimates give equal aggregate weight to each of the 13 groups. Intervals are percentile 95\% intervals from 200,000 cluster bootstrap resamples of those groups.}
\label{tab:counterfactual_fea}
\begin{tabular}{L{5.1cm}rr}
\toprule
Quantity & Estimate & 95\% cluster bootstrap interval \\
\midrule
Diffuse change $R_D$ & $-2.978\%$ & $[-3.152,-2.779]\%$ \\
Localized change $R_L$ & $-3.515\%$ & $[-3.830,-3.197]\%$ \\
$C=(E_L-E_D)/E_B$ & $-0.536\%$ & $[-0.796,-0.281]\%$ \\
$E_L-E_D$ & $-0.788$~MPa & $[-1.510,-0.313]$~MPa \\
\bottomrule
\end{tabular}
\end{table}

Both interventions reduced simulated modulus relative to baseline in all 24 specimens. Localized thinning produced the lower modulus in 19 of 24 cores and in 11 of the 13 group mean contrasts. The primary spatial support contrast was $\widehat C=-0.00536$. The 95\% confidence interval was $[-0.00796,-0.00281]$. The sensitivity estimate from the cluster bootstrap weighted by the number of cores was very similar, $-0.535\%$ with 95\% interval $[-0.816,-0.260]\%$. When each grouping unit was omitted in turn, the equal weight per validation group estimate of $C$ remained negative, ranging from $-0.598\%$ to $-0.457\%$. The secondary exact cluster level sign flip test gave two-sided $p=0.00220$.

The prespecified factor 3 resolution sensitivity also completed all 72 simulations. With equal weight assigned to each validation group, the factor 3 effects were $R_D=-2.806\%$ with 95\% interval $[-2.952,-2.639]\%$, $R_L=-3.312\%$ with interval $[-3.590,-3.032]\%$, and the factor 3 spatial support contrast was $C_3=-0.507\%$ with 95\% interval of $[-0.751,-0.266]\%$. Localized thinning again produced the lower modulus in 19 of 24 cores and in 11 of 13 group means. The sign of $C_i$ was identical between factors two and three for all 24 cores, and the contrasts across cores were almost perfectly correlated ($r=0.9986$). 

Relative to factor 2, the factor 3 contrast was attenuated by only $0.0295$ percentage points, with a paired cluster bootstrap interval of $[0.0059,0.0530]$ percentage points. In contrast, with equal weight assigned to each validation group, the ratio of the baseline modulus at factor 3 to that at factor 2 was 1.086 with 95\% interval $[1.074,1.097]$. Thus, the absolute modulus remained resolution dependent, whereas the localized minus diffuse ordering and its uncertainty were stable across the two prespecified coarse scales.

The mechanical ordering differed from the ordering of global persistence image displacement. Diffuse thinning had the larger baseline to counterfactual distance in both $H_0$ and $H_1$ for all 24 specimens, whereas localized thinning produced the larger stiffness loss in most specimens. Consequently, a global persistence image distance from baseline should not be interpreted as a mechanical damage score. The topological representation is sensitive to the type and spatial support of the perturbation, but these global distances do not identify the mechanical consequences of those changes.

\FloatBarrier
\section{Exploratory $H_0$ score and matched pilot}
\label{app:pilot}

\begin{revision}
\subsection{Signed distance $H_0$ Haufe pattern}

The general Haufe transformation is defined once in Equation~\eqref{eq:haufe_general}. Here it is applied to the centered signed distance $H_0$ persistence image block, giving the distinct pattern $a_T^{\mathrm{SD}}$. If a centered linear fit has score $\widehat y=X\widehat\beta$, Equation~\eqref{eq:haufe_general} can equivalently be written
\[
a^{\mathrm{SD}}
=
\frac{\Sigma_{X,w}\widehat\beta}
{\widehat\beta^{\mathsf T}\Sigma_{X,w}\widehat\beta},
\]
when the denominator is positive. This identity makes explicit why a Haufe pattern combines the fitted coefficient vector with predictor covariance. It remains an association with the fitted score rather than a coefficient vector in the original prediction equation, a causal effect, or a gradient with respect to bone geometry.

The ridge coefficient maps and signed distance $H_0$ Haufe patterns were each normalized to unit Euclidean length before stability comparison. Across the 13 outer fits, median pairwise cosine similarity was 0.956 for coefficients and 0.994 for $a_T^{\mathrm{SD}}$. Fifty-seven persistence image pixels lay in the largest 10\% of mean absolute pattern magnitude and retained their sign in at least 75\% of fits in which they were present. The strongest $5\times5$ region covered births approximately $-754$ to $-581~\mu$m and persistences 506 to 675~$\mu$m, with the same sign in every fit. Since the outer training sets overlap, this is stability across related fits rather than 13 independent replications.
\end{revision}

\begin{figure}[htbp]
\centering
\includegraphics[width=0.76\linewidth]{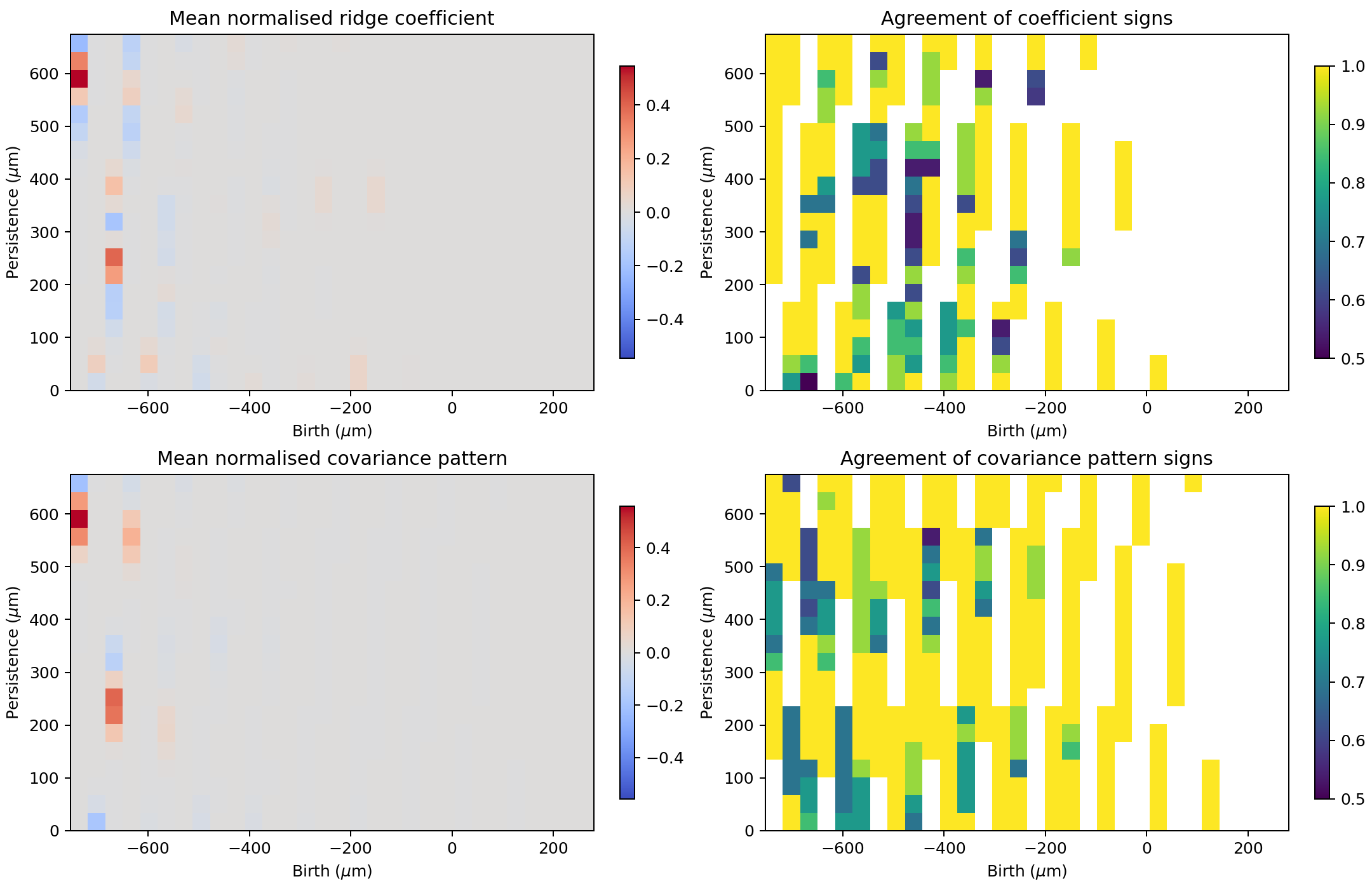}
\caption{\textcolor{revisioncolor}{Original signed distance $H_0$ ridge coefficient and Haufe pattern stability maps. Blank pixels had negligible variance in some training sets. Correlated neighboring pixels summarize regions of the representation, not individual geometric causes.}}
\label{fig:h0_stability}
\end{figure}

\subsection{Candidate matching and pair selection}

A separate single specimen pilot was performed to test whether the direction of the exploratory human $H_0$ association has a corresponding direction in simulated elastic stiffness when the conventional morphometric coordinates are tightly matched. This analysis was designed after the human $H_0$ association had been observed and therefore uses information from that result. It tested the proposed score interpretation without replacing the original predictive or equal removal analyses.

The pilot specimen was selected without using strength, clinical group, persistent homology, or finite element results. Among the 24 baseline ROIs, the coded specimen whose baseline BV/TV was closest to the numerical cohort median was selected, with the coded identifier used only as a deterministic tie breaker. This rule selected QC007. From this baseline, 120 distinct counterfactuals directed toward the bone surface were generated. Every candidate removed exactly 2\% of the baseline bone voxels up to the same integer rounding rule used above and retained the 0.50-mm protected regions at both loading ends. Removal was restricted to a fixed pool of shallow surface voxels ranked by physical Euclidean distance to marrow. The candidate set contained one diffuse candidate spanning the eligible region together with candidates spatially biased toward one axial location, two axial locations, or a three-dimensional focal region. \textcolor{revisioncolor}{Candidate generation did not use strength, the signed distance $H_0$ pattern target, or mechanics.}

BV/TV, Tb.Th, Tb.Sp, Conn.D, and DA were recomputed for every candidate. Because all candidates removed the same integer number of bone voxels from the same ROI, BV/TV was exactly equal across candidates. Let $s_j$ denote the standard deviation of morphometric coordinate $j$ among the 24 human baseline specimens. A candidate pair $(A,B)$ was eligible only when
\[
\frac{|m_{A,j}-m_{B,j}|}{s_j}<0.10
\]
for each $j\in\{\mathrm{Tb.Th},\mathrm{Tb.Sp},\mathrm{Conn.D},\mathrm{DA}\}$. Persistent homology was not used to relax this criterion.

\begin{revision}
Only after the conventional matching set had been determined was the signed distance $H_0$ pattern target applied. The target was recovered from the completed human $H_0$ Haufe pattern analysis rather than re-estimated from the candidate set. Let $\mathcal R$ denote the $5\times5$ persistence image region with the largest mean absolute magnitude of the signed distance patterns normalized to unit norm $\widetilde a_T^{\mathrm{SD}}$. This region covered birth values from approximately $-754$ to $-581~\mu$m and persistence values from 506 to 675~$\mu$m. If $\bar a_r^{\mathrm{SD}}$ denotes the mean normalized signed distance Haufe value at pixel $r\in\mathcal R$, define the regional weights by
\[
w_r^{\mathrm{SD}}
=
\frac{\bar a_r^{\mathrm{SD}}}
{\left(\sum_{q\in\mathcal R}(\bar a_q^{\mathrm{SD}})^2\right)^{1/2}}
\]
and the \emph{regional score} by
\begin{equation}
S_{\mathrm{SD}}(p)
=
\sum_{r\in\mathcal R}w_r^{\mathrm{SD}}p_r.
\label{eq:sd_regional_score}
\end{equation}
The score therefore preserves the signed multivariate pattern within the prespecified region; it is not simply the unweighted persistence image mass in that region.

Among all pairs satisfying the four coordinate morphometric criterion, the selected pair maximized
\[
|S_{\mathrm{SD}}(p_A)-S_{\mathrm{SD}}(p_B)|.
\]
Ties were resolved first by the smaller Euclidean norm of the four standardized morphometric differences and then by candidate identifier. The member with the larger score was designated $H_0$-high and the other $H_0$-low. The pair identities, source image checksums, regional scores, and selection rule were written and hashed before any finite element analysis was performed. The prespecified transfer hypothesis was that, if the positive signed distance pattern observed in the human data has a corresponding ordering of elastic stiffness, the $H_0$-high member would retain the larger modeled mechanical competence.
\end{revision}

The selected pair was evaluated with the same factor 2 fractional-occupancy finite element formulation, material law, boundary conditions, loading, solver, and numerical acceptance gates described above. A factor 3 calculation was then performed as a resolution sensitivity using the same factor 3 construction already specified for the main counterfactual experiment. No pair reselection, morphometric rematching, $H_0$ reweighting, or mechanical retuning was allowed between the two resolutions. For $q\in\{2,3\}$, the mechanical comparison was
\[
\Delta E^{(q)}
=
E_{\mathrm{high}}^{(q)}-E_{\mathrm{low}}^{(q)}.
\]
A positive value would support the prespecified stiffness hypothesis.

\subsection{Selected pair and numerical results}

All 120 candidate structures completed the conventional morphometry and $H_0$ calculations. Of the 7,140 possible unordered candidate pairs, 5,648 satisfied the preliminary Tb.Th, Tb.Sp, and Conn.D criterion, and 5,582 remained eligible after the DA criterion was added. All 120 candidates participated in at least one fully matched pair. The deterministic $H_0$ selection rule chose C000 as $H_0$-low and C109 as $H_0$-high. Their regional scores were 8.4688 and 9.5436, respectively, giving a score separation of 1.0748. C000 was the globally diffuse member of the candidate set, whereas C109 used a focal surface removal pattern.

The selected structures contained exactly 3,027,494 bone voxels each in the same ROI, corresponding to BV/TV of 0.234178. Their remaining morphometric differences were small relative to the variation among the 24 human baselines. Table~\ref{tab:h0_matched_pair} reports the $H_0$-high minus $H_0$-low differences. The Euclidean norm of the four standardized differences was 0.0945, and every coordinate was within the prespecified 0.10 standard deviation bound.

\begin{table}[htbp]
\centering
\caption{Conventional morphometric matching of the pair selected by the $H_0$ regional score. Differences are $H_0$-high minus $H_0$-low. The standardized difference divides by the corresponding standard deviation among the 24 human baseline specimens.}
\label{tab:h0_matched_pair}
\begin{tabular}{L{3.6cm}rr}
\toprule
Feature & Raw difference & Difference / human SD \\
\midrule
BV/TV & 0 & 0 \\
Tb.Th ($\mu$m) & $-2.349$ & $-0.0554$ \\
Tb.Sp ($\mu$m) & $-0.065$ & $-0.0004$ \\
Conn.D (mm$^{-3}$) & $+0.0891$ & $+0.0564$ \\
DA & $-0.00542$ & $-0.0517$ \\
\bottomrule
\end{tabular}
\end{table}

Both variants passed the primary stopping tolerance at both resolutions. Table~\ref{tab:h0_mechanical_pair} gives their moduli. The high minus low differences were $-0.395$~MPa ($-0.385\%$) at factor 2 and $-0.348$~MPa ($-0.303\%$) at factor 3, opposite to the proposed direction. Absolute moduli changed with discretization, but the ordering was preserved.

\begin{table}[htbp]
\centering
\caption{Resolution sensitivity of the morphometry-matched $H_0$ pair. Relative differences use the $H_0$-low modulus as denominator.}
\label{tab:h0_mechanical_pair}
\begin{tabular}{lrrrr}
\toprule
Factor & $E_{\mathrm{low}}$ (MPa) & $E_{\mathrm{high}}$ (MPa) & High--low (MPa) & High--low (\%) \\
\midrule
2 & 102.615 & 102.220 & $-0.395$ & $-0.385$ \\
3 & 114.894 & 114.546 & $-0.348$ & $-0.303$ \\
\bottomrule
\end{tabular}
\end{table}

At factor 2, independently recomputed relative residuals were $9.92\times10^{-7}$ and $9.81\times10^{-7}$ for the low and high structures. Reaction equilibrium errors were below $3.0\times10^{-5}$ and energy inconsistencies below $3\times10^{-8}$. Factor 3 also passed all unchanged numerical gates. These checks support the reported ordering within the stated solver tolerances.

\clearpage
\begingroup
\small
\setlength{\bibsep}{4pt}
\bibliographystyle{unsrtnat}
\bibliography{references}

@article{Cox2026SOI,
  author  = {Cox, Jason M. and Musa, Deeqa M. and Cole, Jacqueline H.},
  title   = {A New Quantitative Index of Trabecular Bone Microstructural Organization and Its Association With Tissue Failure in Human Vertebrae},
  journal = {Journal of Biomechanical Engineering},
  year    = {2026},
  volume  = {148},
  number  = {2},
  pages   = {021010},
  doi     = {10.1115/1.4070322}
}

@article{Liu2011ITS,
  author  = {Liu, X. Sherry and Shane, Elizabeth and McMahon, Donald J. and Guo, X. Edward},
  title   = {Individual Trabecula Segmentation (ITS)-Based Morphological Analysis of Microscale Images of Human Tibial Trabecular Bone at Limited Spatial Resolution},
  journal = {Journal of Bone and Mineral Research},
  year    = {2011},
  volume  = {26},
  number  = {9},
  pages   = {2184--2193},
  doi     = {10.1002/jbmr.420}
}

@misc{Cox2022SOI,
  author = {Cox, Jason M. and Musa, Deeqa M. and Cole, Jacqueline H.},
  title  = {A New Quantitative Index of Trabecular Bone Structural Organization},
  year   = {2022},
  note   = {Orthopaedic Research Society Annual Meeting, Paper No. 1379}
}

@article{Callens2021Minkowski,
  author  = {Callens, Sebastien J. P. and Tourolle n{\'e} Betts, Duncan C. and M{\"u}ller, Ralph and Zadpoor, Amir A.},
  title   = {The local and global geometry of trabecular bone},
  journal = {Acta Biomaterialia},
  year    = {2021},
  volume  = {130},
  pages   = {343--361},
  doi     = {10.1016/j.actbio.2021.06.013}
}

@article{Pothuaud2002,
  author  = {Pothuaud, Laurent and van Rietbergen, Bert and Mosekilde, Lis and Beuf, Olivier and Levitz, Pierre and Benhamou, Claude L. and Majumdar, Sharmila},
  title   = {Combination of topological parameters and bone volume fraction better predicts the mechanical properties of trabecular bone},
  journal = {Journal of Biomechanics},
  year    = {2002},
  volume  = {35},
  number  = {8},
  pages   = {1091--1099},
  doi     = {10.1016/S0021-9290(02)00060-X}
}

@article{Gomberg2000,
  author  = {Gomberg, Bryon R. and Saha, Punam K. and Song, Hee Kwon and Hwang, Scott N. and Wehrli, Felix W.},
  title   = {Topological analysis of trabecular bone MR images},
  journal = {IEEE Transactions on Medical Imaging},
  year    = {2000},
  volume  = {19},
  number  = {3},
  pages   = {166--174},
  doi     = {10.1109/42.845175}
}

@article{vanRietbergen1995,
  author  = {van Rietbergen, B. and Weinans, H. and Huiskes, R. and Odgaard, A.},
  title   = {A new method to determine trabecular bone elastic properties and loading using micromechanical finite-element models},
  journal = {Journal of Biomechanics},
  year    = {1995},
  volume  = {28},
  number  = {1},
  pages   = {69--81},
  doi     = {10.1016/0021-9290(95)80008-5}
}

@article{vanRietbergen1996,
  author  = {van Rietbergen, B. and Weinans, H. and Huiskes, R. and Polman, B. J. W.},
  title   = {Computational strategies for iterative solutions of large FEM applications employing voxel data},
  journal = {International Journal for Numerical Methods in Engineering},
  year    = {1996},
  volume  = {39},
  number  = {16},
  pages   = {2743--2767},
  doi     = {10.1002/(SICI)1097-0207(19960830)39:16<2743::AID-NME974>3.0.CO;2-A}
}

@book{gudhi:urm,
  title = {{GUDHI} User and Reference Manual},
  author = {{The GUDHI Project}},
  publisher = {{GUDHI Editorial Board}},
  year = {2015},
  url = {http://gudhi.gforge.inria.fr/doc/latest/}
}

@article{Zomorodian2004,
  title = {Computing Persistent Homology},
  volume = {33},
  ISSN = {1432-0444},
  DOI = {10.1007/s00454-004-1146-y},
  number = {2},
  journal = {Discrete \& Computational Geometry},
  publisher = {Springer Science and Business Media LLC},
  author = {Zomorodian,  Afra and Carlsson,  Gunnar},
  year = {2004},
  month = Nov,
  pages = {249–274}
}

@misc{Jin_2016, 
 title={The effect of bisphosphonates on bone microstructure and strength}, 
 DOI={10.25560/49791}, 
 author={Jin, Andi}, 
 year={2016}, 
 month={05}
}

@article{CohenSteiner2006,
  title = {Stability of Persistence Diagrams},
  volume = {37},
  ISSN = {1432-0444},
  DOI = {10.1007/s00454-006-1276-5},
  number = {1},
  journal = {Discrete \& Computational Geometry},
  publisher = {Springer Science and Business Media LLC},
  author = {Cohen-Steiner,  David and Edelsbrunner,  Herbert and Harer,  John},
  year = {2006},
  month = dec,
  pages = {103–120}
}

@book{Kaczynski2004,
  title = {Computational Homology},
  ISBN = {9780387215976},
  ISSN = {2196-968X},
  DOI = {10.1007/b97315},
  journal = {Applied Mathematical Sciences},
  publisher = {Springer New York},
  author = {Kaczynski,  Tomasz and Mischaikow,  Konstantin and Mrozek,  Marian},
  year = {2004}
}

@article{ODGAARD1997315,
title = {Three-dimensional methods for quantification of cancellous bone architecture},
journal = {Bone},
volume = {20},
number = {4},
pages = {315-328},
year = {1997},
issn = {8756-3282},
doi = {10.1016/S8756-3282(97)00007-0},
author = {A. Odgaard}
}

@article{ODGAARD1993173,
title = {Quantification of connectivity in cancellous bone, with special emphasis on 3-D reconstructions},
journal = {Bone},
volume = {14},
number = {2},
pages = {173-182},
year = {1993},
issn = {8756-3282},
doi = {10.1016/8756-3282(93)90245-6},
author = {A. Odgaard and H.J.G. Gundersen}
}

@article{SONG2025102857,
title = {Generalized Morse theory of distance functions to surfaces for persistent homology},
journal = {Advances in Applied Mathematics},
volume = {166},
pages = {102857},
year = {2025},
issn = {0196-8858},
doi = {10.1016/j.aam.2025.102857},
author = {Anna Song and Ka Man Yim and Anthea Monod}
}

@article{Hildebrand1997,
author = {Hildebrand, T. and Rüegsegger, P.},
title = {A new method for the model-independent assessment of thickness in three-dimensional images},
journal = {Journal of Microscopy},
volume = {185},
number = {1},
pages = {67-75},
doi = {10.1046/j.1365-2818.1997.1340694.x},
eprint = {https://onlinelibrary.wiley.com/doi/pdf/10.1046/j.1365-2818.1997.1340694.x},
year = {1997}
}

@Article{Pritchard2023,
author={Pritchard, Ysanne
and Sharma, Aikta
and Clarkin, Claire
and Ogden, Helen
and Mahajan, Sumeet
and S{\'a}nchez-Garc{\'i}a, Rub{\'e}n J.},
title={Persistent homology analysis distinguishes pathological bone microstructure in non-linear microscopy images},
journal={Scientific Reports},
year={2023},
month={02},
day={13},
volume={13},
number={1},
pages={2522},
issn={2045-2322},
doi={10.1038/s41598-023-28985-3},
}

@Inbook{Bleile2022,
author="Bleile, Bea
and Garin, Ad{\'e}lie
and Heiss, Teresa
and Maggs, Kelly
and Robins, Vanessa",
title="The Persistent Homology of Dual Digital Image Constructions",
bookTitle="Research in Computational Topology 2",
year="2022",
publisher="Springer International Publishing",
address="Cham",
pages="1--26",
isbn="978-3-030-95519-9",
doi="10.1007/978-3-030-95519-9_1",
}

@article{https://doi.org/10.1111/j.1600-0501.2004.00969.x,
author = {Fanuscu, Mete I. and Chang, Ting-Ling},
title = {Three-dimensional morphometric analysis of human cadaver bone: microstructural data from maxilla and mandible},
journal = {Clinical Oral Implants Research},
volume = {15},
number = {2},
pages = {213-218},
doi = {10.1111/j.1600-0501.2004.00969.x},
year = {2004}
}

@article{vanderWalt2014,
  title = {scikit-image: image processing in Python},
  volume = {2},
  ISSN = {2167-8359},
  DOI = {10.7717/peerj.453},
  journal = {PeerJ},
  publisher = {PeerJ},
  author = {van der Walt,  Stéfan and Sch\"{o}nberger,  Johannes L. and Nunez-Iglesias,  Juan and Boulogne,  Fran\c{c}ois and Warner,  Joshua D. and Yager,  Neil and Gouillart,  Emmanuelle and Yu,  Tony},
  year = {2014},
  month = jun,
  pages = {e453}
}

@ARTICLE{Otsu1979,
  author={Otsu, Nobuyuki},
  journal={IEEE Transactions on Systems, Man, and Cybernetics}, 
  title={A Threshold Selection Method from Gray-Level Histograms}, 
  year={1979},
  volume={9},
  number={1},
  pages={62-66},
  doi={10.1109/TSMC.1979.4310076}}

@article{JMLR:v18:16-337,
  author  = {Henry Adams and Tegan Emerson and Michael Kirby and Rachel Neville and Chris Peterson and Patrick Shipman and Sofya Chepushtanova and Eric Hanson and Francis Motta and Lori Ziegelmeier},
  title   = {Persistence Images: A Stable Vector Representation of Persistent Homology},
  journal = {Journal of Machine Learning Research},
  year    = {2017},
  volume  = {18},
  number  = {8},
  pages   = {1--35},
}

@Article{BoneJ2021,
AUTHOR = { Domander, R and Felder, AA and Doube, M},
TITLE = {BoneJ2 - refactoring established research software [version 2; peer review: 3 approved]
},
JOURNAL = {Wellcome Open Research},
VOLUME = {6},
YEAR = {2021},
NUMBER = {37},
DOI = {10.12688/wellcomeopenres.16619.2}
}

@article{imagej2021,
author = {Schroeder, Alexandra B. and Dobson, Ellen T. A. and Rueden, Curtis T. and Tomancak, Pavel and Jug, Florian and Eliceiri, Kevin W.},
title = {The ImageJ ecosystem: Open-source software for image visualization, processing, and analysis},
journal = {Protein Science},
volume = {30},
number = {1},
pages = {234-249},
doi = {10.1002/pro.3993},
year = {2021}
}

@article{HoerlKennard1970,
  author = {Hoerl, Arthur E. and Kennard, Robert W.},
  title = {Ridge Regression: Biased Estimation for Nonorthogonal Problems},
  journal = {Technometrics},
  year = {1970},
  volume = {12},
  number = {1},
  pages = {55--67},
  doi = {10.1080/00401706.1970.10488634}
}

@article{Haufe2014,
  author = {Haufe, Stefan and Meinecke, Frank and Gorgen, Kai and Dahne, Sven and Haynes, John-Dylan and Blankertz, Benjamin and Biessmann, Felix},
  title = {On the Interpretation of Weight Vectors of Linear Models in Multivariate Neuroimaging},
  journal = {NeuroImage},
  year = {2014},
  volume = {87},
  pages = {96--110},
  doi = {10.1016/j.neuroimage.2013.10.067}
}

@article{Jin2017,
  author = {Jin, A. and Cobb, J. and Hansen, U. and Bhattacharya, R. and Reinhard, C. and Vo, N. and Atwood, R. and Li, J. and Karunaratne, A. and Wiles, C. and Abel, R.},
  title = {The effect of long-term bisphosphonate therapy on trabecular bone strength and microcrack density},
  journal = {Bone \& Joint Research},
  year = {2017},
  volume = {6},
  number = {10},
  pages = {602--609},
  doi = {10.1302/2046-3758.610.BJR-2016-0321.R1}
}

@article{Chevalier2007,
  author = {Chevalier, Yan and Pahr, Dieter and Allmer, Helga and Charlebois, Mathieu and Zysset, Philippe},
  title = {Validation of a voxel-based FE method for prediction of the uniaxial apparent modulus of human trabecular bone using macroscopic mechanical tests and nanoindentation},
  journal = {Journal of Biomechanics},
  year = {2007},
  volume = {40},
  number = {15},
  pages = {3333--3340},
  doi = {10.1016/j.jbiomech.2007.05.004}
}

@misc{Topolnicki2026Directional,
  author = {Topolnicki, Rafa\l{} and Bogdan, Micha\l{} and Malinowski, Jakub and Naskrecki, Bartosz and Hara{\'n}czyk, Maciej and D\l{}otko, Pawe\l{}},
  title = {Direction-aware topological descriptors for elastic stiffness tensor prediction in porous materials},
  year = {2026},
  eprint = {2604.08105},
  archivePrefix = {arXiv},
  primaryClass = {physics.comp-ph},
  doi = {10.48550/arXiv.2604.08105}
}

@inproceedings{Kornblith2019CKA,
  title = {Similarity of Neural Network Representations Revisited},
  author = {Kornblith, Simon and Norouzi, Mohammad and Lee, Honglak and Hinton, Geoffrey},
  booktitle = {Proceedings of the 36th International Conference on Machine Learning},
  series = {Proceedings of Machine Learning Research},
  volume = {97},
  pages = {3519--3529},
  year = {2019},
  publisher = {PMLR},
  url = {https://proceedings.mlr.press/v97/kornblith19a.html}
}
\endgroup
\end{document}